\documentstyle[12pt]{article}
\newlength{\extraspace}
\newlength{\extraspaces}
\catcode`\@=11
\def\numberbysection{\@addtoreset{equation}{section}
\def\theequation{\arabic{section}.\arabic{equation}}}

\begin{document}
\addtolength{\baselineskip}{.7mm}
\thispagestyle{empty}
\begin{flushright}
TIT/HEP--317 \\
UCLA/96/TEP/11 \\
{\tt hep-th/9603206} \\
March, 1996 
\end{flushright}
\vspace{2mm}
%
%
\begin{center}
{\large{\bf   Gauge Symmetry Breaking through Soft Masses}}
\vspace{5pt}
{\large{\bf   in Supersymmetric Gauge Theories }} \\[15mm]
{\sc Eric D'Hoker}\footnote{
\tt e-mail: dhoker@physics.ucla.edu}   \\[2mm]
{\it Department of Physics and Astronomy, \\
 University of California at Los Angeles\\
Los Angeles, CA 90024, USA} \\[4mm]
{\sc Yukihiro Mimura}\footnote{
\tt e-mail: mim@th.phys.titech.ac.jp}  \hspace{2mm} 
 and  \hspace{2mm}
{\sc Norisuke Sakai}\footnote{
\tt e-mail: nsakai@th.phys.titech.ac.jp}   \\[2mm]
{\it Department of Physics, Tokyo Institute of Technology \\
Oh-okayama, Meguro, Tokyo 152, Japan}  \\[10mm]
{\bf Abstract}\\[5mm]
{\parbox{14cm}{\hspace{5mm}
We analyze the effects of soft supersymmetry breaking terms on $N=1$ 
supersymmetric QCD with $N_f$ flavors and color gauge group $SU(N_c)$.  
The mass squared of some squarks may be negative, as long as vacuum 
stability is ensured by a simple mass inequality. 
For $N_f<N_c$, we include the dynamics of the 
non-perturbative superpotential and use the original (s)quark and 
gauge fields, while for $N_f>N_c+1$, we formulate the dynamics in 
terms of dual (s)quarks and a dual gauge group $SU(N_f-N_c)$. 
The presence of negative squark mass squared terms leads to spontaneous 
breakdown of flavor and color symmetry.
We determine this breaking pattern, derive the spectrum, and argue that 
the masses vary smoothly as one crosses from the Higgs phase into the 
confining phase. 
 }}
\end{center}
\vfill
\newpage
\setcounter{section}{0}
\setcounter{equation}{0}
\setcounter{footnote}{0}
\def\theequation{\arabic{section}.\arabic{equation}}
%
%
%
\vspace{7mm}
\pagebreak[3]
\addtocounter{section}{1}
\setcounter{equation}{0}
\setcounter{subsection}{0}
\setcounter{footnote}{0}
\begin{center}
{\large {\bf \thesection. Introduction}}
\end{center}
\nopagebreak
\medskip
\nopagebreak
\hspace{3mm}
The understanding of the dynamics of supersymmetric gauge theories 
has vastly improved through a combined use of the kinematical 
constraints of holomorphy and the dynamical assumption of duality. 
Whereas a number of exact results are known in theories with 
extended supersymmetry \cite{SeWi}, 
a reliable qualitative picture has emerged in theories with 
simple ($N=1$) supersymmetry as well \cite{VeYa} -- \cite{InLeSe}. 
Properties such as confinement and dynamical chiral symmetry 
breaking are found to interplay in a variety of different ways, 
depending on the numbers of colors $N_c$ and 
flavors $N_f$ of quarks and squarks. 

Seiberg has proposed that these features are not restricted to $N=1 $
supersymmetric gauge theories, but should survive -- at least in a 
qualitative way -- to non-supersymmetric gauge theories. 
The models in which this proposal is perhaps most easily verified 
are those where supersymmetry is spontaneously 
broken (through the introduction of an additional sector of 
fields \cite{FaIl} -- \cite{Oraifeartaigh}) or 
those where soft, explicit supersymmetry breaking terms are added as a 
perturbation on the gauge dynamics \cite{GiGr}. 
The latter scheme of supersymmetry breaking was used in the original 
proposal of supersymmetric grand unified theories \cite{Sakai}, 
\cite{DiGe} and provides a general framework for 
the usual formulation of the Minimal Supersymmetric 
Standard Model \cite{Nilles}. 

In a series of papers, it was argued that the addition of 
perturbative, soft 
supersymmetry breaking mass terms, with $m^2 \geq 0$, essentially 
preserves the 
qualitative picture of the dynamics derived for $N=1$ supersymmetric 
QCD (SQCD) \cite{AhSoPe}. 
An effective low energy theory is used
in terms of color singlet meson and (for $N_f\geq N_c$) 
baryon fields, appropriate for the confining phase, and 
the effects of the non-perturbative 
superpotential of Affleck, Dine and Seiberg \cite{AfDiSe} are included. 
It was shown that flavor symmetry is dynamically broken from 
$SU(N_f) \times SU(N_f)$ down to $SU(N_f)$, just as in standard 
non-supersymmetric QCD. 
Indications are that, mostly, the dynamics admits a smooth transition 
to that of standard QCD, corresponding to the limit of large soft
breaking  mass terms. 
{}For special arrangements of $N_c$ and $N_f$, however, phase 
transitions may occur as the mass is increased, a possibility 
considered by \cite{PrWe}.

In the present paper, we investigate $N=1$ supersymmetric QCD, again 
with soft supersymmetry breaking mass terms added, but this time with 
$m^2<0$  for at least some of the squark fields. 
{}For general values of the soft 
supersymmetry breaking mass terms with $m^2<0$, the Hamiltonian will 
become unbounded from below, thus destabilizing the entire theory.  
{}For certain simple ranges of the masses, however, we show that 
stable vacua exist through a balance between the soft supersymmetry 
breaking mass terms, the quartic $D^2$ term 
for the squark fields and ( for $N_f < N_c$ ) the 
non-perturbative effective potential. 
We argue that in these solutions, flavor as well as color symmetry 
are spontaneously 
broken through the vacuum expectation value of the squarks. 
We consider the most general soft supersymmetry breaking terms 
respecting $R$-symmetry, for simplicity. 

{}For the analysis of spontaneous flavor and color symmetry breakdown 
when $N_f<N_c$, we formulate the dynamics in terms of the fundamental
quark superfields instead of in terms of meson superfields, in 
contrast with the analysis in ref.\cite{AhSoPe}. 
This choice appears more natural when dealing with the theory in the
Higgs phase, 
rather than in the confining phase. 
In fact, the original calculation of the nonperturbative superpotential 
was justified precisely by considering gauge symmetry breakdown due to 
the vacuum expectation values of squark fields \cite{AfDiSe}. 

{}For small values of the soft supersymmetry breaking mass terms, we 
expect the confining and Higgs phases to be smoothly connected 
to one another, with matching low energy spectra. 
We find that this principle of 
complementarity between the confining and Higgs phases \cite{FrSh}
can indeed be satisfied in these theories. To do so however, it appears 
to be necessary to consider the most general low energy effective action 
(consistent with internal symmetries and supersymmetry) in the confining 
phase. This action must include terms of higher order than was 
originally advocated in ref.\cite{AhSoPe}.

{}For large values of the soft supersymmetry breaking mass terms, we 
expect the semi-classical spectrum for the Higgs phase, derived in 
terms of the fundamental quark superfields, to remain reliable.
Thus, we shall calculate the 
semi-classical spectrum for this model for all ranges of soft 
breaking mass terms. 
{}For $N_f > N_c +1 $, we will use the dual variables of 
ref.\cite{Seiberg}, 
which are more appropriate to describe the spontaneous symmetry 
breakdown of flavor and (dual) color symmetry. 

{}Finally, we note that there is a simple extension of standard SQCD, 
obtained by gauging also an additional anomaly-free $U(1)_X$ symmetry, 
and adding a Fayet-Illiopoulos $D$-term \cite{FaIl} for the 
corresponding gauge multiplet. (The simplest case would be where this 
$U(1)_X$ is just baryon number symmetry, but any anomaly-free 
$U(1)_X$ would do.) 
No explicit supersymmetry breaking terms are added; instead, a 
supersymmetric mass term is included in the superpotential, which 
stabilizes the vacuum. 
In this model, supersymmetry is broken spontaneously, and 
soft mass terms with $m^2<0$ automatically arise from 
the Fayet-Illiopoulos $D$-term.  
This model provides an economical realization of some of the
effects of soft supersymmetry breaking mass terms generated directly by 
spontaneous breakdown of  supersymmetry. 
It will be discussed in a companion paper.
A different model has been analyzed which obtains the soft breaking 
terms from the spontaneous breakdown of supersymmetry \cite{EvHsSc}. 

The remainder of the paper is organized as follows. 
In Sect. 2, we analyze the case where $N_f < N_c$, and propose 
the use of the original squark fields as 
relevant effective degrees of freedom when some of the soft mass 
terms have $m^2<0$. 
Including kinetic terms, gauge couplings, soft breaking 
mass terms and 
the non-perturbative effective potential of ref.\cite{AfDiSe}, 
we find the ranges of stability for the soft breaking mass terms, 
and the patterns of flavor and color symmetry breakdown. 
In Sect. 3, we analyze the case of supersymmetric QCD with 
$
N_f > N_c+1 $, with the help of the duality correspondence 
of Seiberg. 
Here, we add mass terms not to the original squark fields of the 
theory, but rather to the dual squark fields, which carry the 
dual color quantum numbers of $SU(N_f -N_c)$. 
In Sect. 4, 
we point out the difficulties encountered with 
our approach for $N_f = N_c$ and $N_f = N_c +1$ and 
discuss the region of validity of our analysis. 

%
%
\vspace{7mm}
\pagebreak[3]
\addtocounter{section}{1}
\setcounter{equation}{0}
\setcounter{subsection}{0}
\setcounter{footnote}{0}
\begin{center}
{\large {\bf \thesection. Dynamics for $N_f < N_c$
}}
\end{center}
\nopagebreak
\medskip
\nopagebreak
\hspace{3mm}
In this Section, we shall consider supersymmetric Yang-Mills theory 
with gauge group $SU(N_c)$ and $N_f$ flavors of squarks and quarks 
(with $N_f < N_c$), transforming under the representation 
$N_c\oplus \bar  N_c$ of $SU(N_c)$. This 
theory is the natural supersymmetric extension of QCD, 
and will be referred to as SQCD. 
The corresponding chiral superfields 
\begin{equation}
\hat Q_a{}^i \; 
\qquad 
\hat {\bar Q} _i{}^a
\qquad \qquad
a=1, \cdots, N_c;
\qquad
i=1, \cdots , N_f\; ,
\label{eq:one}
\end{equation} 
contain the squark fields $Q$ and $\bar Q$ and the left-handed quark 
fields
$\psi _Q$ and $\psi _{\bar Q}$ respectively. There is a natural color 
singlet
meson chiral superfield $\hat T$, defined by
\begin{equation}
\hat T_i{}^j = \hat {\bar Q}_i{}^a \hat Q_a{}^j
\label{eq:two}
\end{equation}
with scalar components $T_i{}^j$. 
Superfields are denoted by a cap on the scalar components. 

As a starting point we consider classical massless SQCD 
whose Lagrangian ${\cal L}_0$ is 
determined by $SU(N_c)$ gauge invariance, 
by requiring that the 
superpotential for the quark superfields vanish identically :
\begin{equation}
{\cal L}_0 = \int d^4 \theta {\rm \, tr} \{
\hat Q ^{\dagger}  e^{2g\hat V}\hat Q + 
\hat {\bar Q}  e^{-2g\hat V}\hat{\bar Q} ^{\dagger} \} 
+ \frac{1}{2} \int  d^2\theta ~{\rm \, tr} W W 
+ \frac{1}{2} \int  d^2\bar \theta ~{\rm \, tr} \bar W \bar  W 
\label{eq:three}
\end{equation}
This theory has a global symmetry, 
$G_f=SU(N_f)_Q\times SU(N_f)_{\bar Q} \times
U(1)_B\times U(1)_R$, under which the (bosonic or left-handed fermionic)
component fields transform as
\begin{eqnarray}
Q      \to & U_c ~Q ~U_Q^\dagger  \qquad \qquad U_Q \in & SU(N_f)_Q 
\nonumber \\
\bar Q \to & U_{\bar Q} ~\bar Q ~U_c^\dagger \qquad \qquad U_{\bar Q} 
\in &
        SU(N_f)_{\bar Q} 
\label{eq:four} \\
T \to & U_{\bar Q}~ T ~U_Q ^\dagger \qquad\qquad U_c \in & SU(N_c) \; , 
\nonumber 
\end{eqnarray}
with baryon number charges 1, -1 and 0 respectively, and $R$-charges 
given by
\begin{equation}
Q, ~\bar Q ~ : ~ 1-N_c/N_f 
\qquad \qquad
\psi _Q, ~ \psi _{\bar Q} ~ : ~ -N_c/N_f
\label{eq:five}
\end{equation}
A further classical axial baryon number $U(1)_{AB}$ symmetry 
suffers a color anomaly and is absent at the quantum level.
\par
%
%
%
%
Exact nonperturbative results in supersymmetric gauge theories 
can be given for the $F$-type term which is a chiral superspace 
integral of a superpotential $W_{NP}$ \cite{AfDiSe} of 
the quark superfields $\hat Q$ and $\hat {\bar Q}$ given as follows
\begin{equation}
\int d^2 \theta W_{NP}(\hat Q, \hat {\bar Q}) = (N_c-N_f) \Lambda
^{3+2N_{f}/(N_c-N_f)}
\int d^2\theta ~({\rm det} \hat {\bar Q}\hat Q ) ^{-1/(N_c - N_f)} 
\label{eq:seven}
\end{equation}
The parameter $\Lambda$ has dimensions of mass and characterizes the 
strength of the non-perturbative effects. Henceforth, we shall set this 
parameter to $\Lambda =1$ to simplify notations; its dependence is easily
recovered on dimensional grounds. 
Nonperturbative corrections are possible for $D$-type terms such as 
kinetic terms, but the exact results are not available since  
holomorphy places no direct restrictions on them \cite{Seiberg}. 

%
\vspace{5mm}
\pagebreak[3]
\addtocounter{section}{0}
\addtocounter{subsection}{1}
\setcounter{footnote}{0}
\begin{center}
{\bf \thesubsection. Soft Supersymmetry Breaking Mass Terms}
\end{center}
\nopagebreak
\medskip
\nopagebreak
\hspace{2mm}
We choose to break supersymmetry explicitly, by adding to the 
Lagrangian ${\cal L}_0$ soft supersymmetry breaking terms for 
the quark supermultiplet. 
In general, soft breaking operators are defined as those which do not 
induce quadratic  divergences to any order in perturbation theory 
(apart from vacuum diagrams).  The only possible soft breaking 
operators can be summarized as in \cite{GiGr}~: 
quadratic and cubic terms involving scalar fields 
with the same chirality; quadratic terms between scalar fields with 
opposite chirality (scalar masses); and gaugino masses. 
With the field contents of our model, 
the color gauge invariance forbids the cubic terms. 
The quadratic terms with the same chirality is forbidden if we want to 
maintain the flavor symmetry $SU(N_f)_Q\times SU(N_f)_{\bar Q}$. 
The gaugino masses violate the $U(1)_R$ symmetry. 
On the other hand squark masses that are separately equal on $Q$ 
and on $\bar Q$ will preserve the entire  flavor group invariance 
$SU(N_f)_Q\times SU(N_f)_{\bar Q} \times U(1)_B\times U(1)_R$. 
Therefore we shall assume in this paper that squark masses are the 
only soft supersymmetry breaking terms. 

Besides the above soft supersymmetry breaking operators, there are also 
the supersymmetry preserving  masses for quark and antiquark 
superfields, which are evidently soft. 
However, their presence always explicitly breaks flavor symmetry. 
For example, if a common supersymmetric mass is given to all quarks, 
the flavor symmetry group   
$SU(N_f)_Q\times SU(N_f)_{\bar Q} \times U(1)_B\times U(1)_R$  will be 
broken down to $SU(N_f)\times U(1)_B\times U(1)_R$. 

{}For the sake of simplicity, we shall add to the  Lagrangian only soft 
supersymmetry breaking squark mass terms and  neglect effects due to 
gaugino masses and supersymmetric flavor masses.   
Generic mass squared for squark and antisquarks are given by matrices 
$M^2_Q$ and $M^2_{\bar Q}$ 
\begin{equation}
{\cal L} _{sb} = - \{ {\rm \, tr} Q M_{Q}^{2} Q^\dagger  
+ {\rm \, tr} \bar Q^\dagger M^2_{\bar Q} \bar Q 
 \}
\label{eq:six}
\end{equation}
As we remarked above, when $M^2_Q$ and $M^2_{\bar Q}$ 
are proportional to the identity matrix, 
the global flavor symmetry is unchanged~: $SU(N_f)_Q \times
SU(N_f)_{\bar Q}\times U(1)_B
\times U(1)_R$.

The soft supersymmetry breaking masses can be expressed in terms of 
spurion superfields $\hat \eta$ and $\hat {\bar \eta}$ in superspace 
formalism \cite{GiGr} 
\begin{equation}
{\cal L} _{sb} = - \int d^4\theta 
\left( 
 {\rm \, tr} \hat Q \hat \eta \hat Q^\dagger  
+ {\rm \, tr} \hat {\bar Q}^\dagger 
\hat {\bar \eta} \hat {\bar Q} 
\right)
\label{eq:susybrspurion}
\end{equation}
\begin{equation}
\hat \eta = M_{Q}^{2} \theta^2 \bar \theta^2, 
\qquad 
\hat {\bar \eta} = M^2_{\bar Q} \theta^2 \bar \theta^2 
\label{eq:spurion}
\end{equation}
This expression can be regarded as a spurion 
insertion to the kinetic term of the superfields $\hat Q$ and 
$\hat {\bar Q}$. 
As long as we insist on the spurion $\hat \eta$ and $\hat {\bar \eta}$ 
as the only source of supersymmetry breaking, we do not have any 
other possible supersymmetry breaking terms in the entire Lagrangian. 
This is because all the other terms including the nonperturbative 
superpotential (\ref{eq:seven}) do not allow for any dependence on the 
spurion superfields $\hat \eta, \hat {\bar \eta}$, which are not 
chiral. 
If one assumes that chiral spurion superfields $\hat \rho=m\theta^2$ 
are also present as a source of supersymmetry breaking, 
one finds supersymmetry breaking terms of the F-type arising from the 
superpotential, such as the nonperturbative one (\ref{eq:seven}). 
We will only use the spurion of the general superfields 
$\hat \eta, \hat {\bar \eta}$ in eq.(\ref{eq:spurion}) in this work.

When both $M^2_Q$ and $M^2_{\bar Q}$ are positive definite, the vacuum
expectation value of the fields $Q$ and $\bar Q$ will vanish, and 
no spontaneous flavor symmetry breakdown is expected to occur 
(neglecting possible effects due to the nonperturbative potential
in eq.(\ref{eq:seven})
).  
Color confinement is expected to take place and the physical spectrum 
should  consist solely of color-singlet hadrons. 
This is the so-called  confining phase. 
It is appropriate to reformulate the low energy dynamics of the theory 
in terms of color singlet fields only. 
This is precisely the approach followed by refs. \cite{Seiberg}, 
\cite{AhSoPe} which formulate the effective low energy dynamics of the 
theory in terms of 
\footnote{For the case of $N_f<N_c$ under consideration, baryon fields 
do not appear.}
the meson fields $T_i{}^j$, introduced in eq.(\ref{eq:two}). 

When either $M^2_Q$ or $M^2_{\bar Q}$ is not positive definite, 
we expect the pattern of symmetry breaking to be
substantially  different. Global flavor symmetry should be 
spontaneously broken,
and $Q$ and/or $\bar Q$ should acquire non-vanishing vacuum 
expectation values.
These non-zero vacuum expectation values, in turn, are expected 
to break color
$SU(N_c)$ and give mass to some of the gauge particles through 
the Higgs
mechanism. This is the so-called Higgs phase.  Exactly how much is left
unbroken of the color symmetry depends upon the number of squark 
masses squared
that are taken to be negative. We shall see that only a limited number
of negative values can occur if the Hamiltonian is to remain bounded
from below. 

According to standard lore, (originally derived from lattice gauge 
theory) the confining and Higgs phases are smoothly connected to one 
another in at least some region of parameter space \cite{FrSh}. 
There should be a one to one correspondence between the 
observables in both phases, suggesting that -- in principle -- 
color singlet meson fields could still be used to describe the dynamics 
of the Higgs phase.  In practice, however, a formulation in terms of
colored fields appears more suitable instead. 
Indeed, physical free quarks and certain massive gauge bosons 
are expected to appear in the low energy spectrum, and  
it is unclear how to represent these degrees of freedom in 
terms of meson variables. 
Thus, we shall use the original squark $Q,~\bar Q$, 
quark $\psi _Q, ~ \psi _{\bar Q}$, and gauge boson and fermion fields 
as physical variables at low energy. 
%
%
\vspace{5mm}
\pagebreak[3]
\addtocounter{section}{0}
\addtocounter{subsection}{1}
\setcounter{footnote}{0}
\begin{center}
{\bf \thesubsection. Vacuum Stability }
\end{center}
\nopagebreak
\medskip
\nopagebreak
\hspace{2mm}
We now have all the necessary ingredients to construct the potential 
of the squark fields : the effective non-perturbative contribution to 
the superpotential of~(\ref{eq:seven}), the soft supersymmetry breaking 
mass terms of (\ref{eq:six}) and finally, the square of the color 
$D$-term arising from the gauge couplings of the $\hat {\bar Q}$ and 
$\hat Q$ superfields, as given by (\ref{eq:three}). 
These contributions are easily worked out in terms of the 
squark fields, and we have 
\footnote{Notice that the normalization of the kinetic term, i.e. 
the K\"ahler potential in $\hat Q$ and $\hat {\bar Q}$ is canonically 
equal to 1, so that no new normalization parameter need to be 
introduced here, in opposition to \cite{AhSoPe}.}
\begin{eqnarray}
V  
&= & {\rm \, tr}  \bigl \{ 
Q (Q^\dagger \bar Q ^\dagger \bar Q Q) ^{-1} Q^\dagger 
+ \bar Q ^\dagger ( \bar Q Q  Q ^\dagger \bar Q ^\dagger ) ^{-1} 
 \bar Q \bigr \} ~ \bigl |{\rm det}  \bar Q Q \bigr |^{-2 /(N_c -N_f)}
\nonumber \\ 
& & + {g^2 \over 2} \bigl ( {\rm \, tr} Q^\dagger t^a Q - {\rm \, tr} 
\bar Q t^a \bar Q
^\dagger)^2 + \bigl 
\{ {\rm \, tr} Q M_{Q}^{2} Q^\dagger  + {\rm \, tr} \bar Q^\dagger 
M^2_{\bar Q} \bar Q 
 \}
\label{eq:eight}
\end{eqnarray}
where the $SU(N_c)$ generators $t^a$ are normalized as 
${\rm \, tr} t^a t^b = {1 \over 2}\delta^{ab}$. 

{}For zero soft supersymmetry breaking masses $M^2_{Q}$ and 
$M^2_{\bar Q}$, we recover standard massless SQCD. This theory 
possess no physical vacuum state, as can be  seen by analyzing 
the full potential of (\ref{eq:eight}) for $M^2_{Q}=M^2_{\bar Q}=0$.  
The potential diverges whenever an eigenvalue of $\bar Q Q$ tends to 
zero, thus driving its minimum  away from $Q=0$ and $\bar Q=0$. 
Since both the quartic and the non-perturbative terms are positive, we 
obtain the absolute minimum configuration, with zero value of the 
potential, by setting $Q= \bar Q^\dagger$, (thus cancelling the quartic 
term) and letting $Q=\bar Q ^\dagger \to \infty$. This is a  runaway 
solution and there is no physical vacuum state.

{}For positive values of squark mass squared, the above runaway solution 
is stabilized and a physical vacuum for SQCD is generated. 

{}For generic matrices $M^2_Q$ and $M^2_{\bar Q}$, (in particular, with 
negative eigenvalues) the runaway directions can turn into directions 
along which the potential is unbounded from below. 
This is now possible, because the soft mass terms break supersymmetry 
explicitly, and the Hamiltonian need not be positive any longer. 
Restrictions on the soft mass terms guaranteeing that the potential be 
bounded from below can be obtained as follows. 
We use global flavor transformations to render the mass squared 
matrices $M_Q^2$ and $M_{\bar Q}^2$ diagonal in flavor, and denote their
eigenvalues by $m_{Q_i}^2$ and $m_{\bar Q_i}^2$ respectively.  Along the 
special directions $Q_i=\bar Q^{\dagger}_j$,  the quartic potential 
vanishes identically and the non-perturbative potential tends  to zero 
as $Q_i=\bar Q_j^{\dagger} \to \infty$.  If the potential is to be 
bounded from below, the remaining soft 
supersymmetry breaking mass terms in the potential must satisfy 
\begin{equation}
m_{Q_i}^2 + m_{\bar Q_j}^2 \geq 0 
\label{eq:nine}
\end{equation}
for any pair of $i, j=1,\cdots N_f$. 
It is easy to see that this condition is also sufficient, 
as we shall establish later by explicit calculation of the potential in a
convenient  set of coordinates for the fields $Q$ and $\bar Q$.

Requiring in addition that the theory possess a well-defined ground 
state, leads to the condition above, but with strict inequalities, 
which is assumed henceforth.

The addition of supersymmetric quark mass terms would produce a
stabilizing effect analogous to that of the positive soft mass squared 
terms. 
The pattern of gauge symmetry breakdown due to the supersymmetric 
mass terms has been analyzed in ref.\cite{InLeSe}. 
If we add the supersymmetric mass terms, we can safely take the limit 
of vanishing soft mass terms without spoiling stability. 
Our results in this limit smoothly connect to the analysis in 
ref.\cite{InLeSe}. 

\vspace{5mm}
\pagebreak[3]
\addtocounter{section}{0}
\addtocounter{subsection}{1}
\setcounter{footnote}{0}
\begin{center}
{\bf \thesubsection. Description of Vacuum Configurations }
\end{center}
\nopagebreak
\medskip
\nopagebreak
\hspace{2mm}

{}For simplicity, we shall explicitly analyze only the case where the 
mass squared for all $Q$'s and $\bar Q$'s are equal to $-m_{Q}^2$ and 
$m_{\bar Q}^2$ respectively, thus preserving the entire global symmetry 
$SU(N_f)_Q\times SU(N_f)_{\bar Q} \times U(1)_B \times U(1)_R$. 
The vacuum configuration is assumed to be Poincar\'e invariant and
so that the values of $Q$ and $\bar Q$ in the vacuum are space-time 
independent. We shall determine these expectation values at
the semi-classical level. 

By making a global $SU(N_c)\times SU(N_f)_Q \times U(1)_B$ 
transformation on $Q$, and the remaining 
$SU(N_f)_{\bar Q}\times U(1)_R$ transformation on $\bar Q$, 
we can always rotate the vacuum expectation values of $Q$ and 
$\bar Q$ to the following arrangement 
\begin{eqnarray}
Q
&= 
\left ( \begin{array}{c} Q_{(1)} \\ 0 \\ \end{array} \right ) 
\qquad \qquad \qquad \quad \
(Q_{(1)})_i{}^j  
& =
Q_i \delta _i {}^j  
\nonumber \\
\bar Q 
&= 
\left ( \begin{array}{cc} \bar Q_{(\bar 1)} U^\dagger 
& \bar Q_{(\bar 2)} \\
\end{array} \right )
\qquad \qquad
(\bar Q _{(\bar 1)}) _i{}^j 
&=
\bar Q_i \delta _i{}^j 
\label{eq:ten} 
\end{eqnarray}
The eigenvalues $Q_i$ and $\bar Q_i$, $i=1,\cdots, N_f$, can be chosen 
to be real and positive. 
The matrix $U\in SU(N_f)$ arises from the fact that color must
act on both $Q$ and $\bar Q^\dagger$ in the same way, so that 
independent color rotations can be made only on $Q$. 
The matrix $\bar Q_{(\bar 2)}$ is a general
$N_f \times (N_c-N_f)$ complex matrix. 
This choice of variables is natural and has the advantage of considerably
simplifying the non-perturbative potential contribution, especially the
determinant. 

To simplify the quartic contribution as well, it will be 
helpful to split up the generators $t^p$ of $SU(N_c)$ according to 
the splitting inherent in eq.(\ref{eq:ten}). 
We separate $t^p$ into diagonal generators $H^p$, $p=1,\cdots
N_c-1$, (of the Cartan subalgebra) and off diagonal (or root) generators 
$E^p$. The diagonal generators can be further separated as follows
\begin{eqnarray}
H_{(0)} 
= 
&  \left ( 
\begin{array}{cc} I_{{\scriptscriptstyle N_{f}}} C_{1} & 0 \\ 0 &
-I_{{\scriptscriptstyle N_{c} - N_{f}}}C_{2}
\end{array}
\right),  
\qquad
\begin{array}{c} C_{1} 
= {\scriptstyle \sqrt{(N_{c}-N_{f})/2 N_f N_c }} \\
                 C_{2} 
= {\scriptstyle \sqrt{N_{f}/2N_{c} (N_{c}-N_{f})}},
\end{array}
\nonumber \\ 
& \nonumber  \\
H^p_{(1)} 
=& \left (\begin{array}{cc} h^p_{(1)} & 0 \\ 0 & 0 \\
\end{array}\right),
\qquad \qquad \qquad \qquad
H^q_{(2)} 
= \left (\begin{array}{cc} 0 & 0 \\ 0 &  h^q_{(2)}\\ \end{array}\right). 
\label{eq:eleven}
\end{eqnarray}
Here, $I_n$ denotes the $n\times n$ identity matrix; the matrices
$h^p_{(1)}$ and $h^q_{(2)}$ are of dimension $N_f\times N_f$ and 
$(N_c-N_f)\times (N_c-N_f)$ respectively, and are given by
\begin{equation}
(h^p)_{i}{}^{j} =    \frac{\delta _{i}{}^{j}}{\sqrt{2p(p+1)}}
\left
\{
\begin{array}{ll}
1  & \mbox{for $1 \leq i \leq p$} \\
-p & \mbox{for $i=p+1$} \\
0  & \mbox{for $i \geq p + 2$} \\
\end{array} \right.
\label{eq:twelve}
\end{equation}
for $h_{(1)}^p$, $p= 1, \cdots, N_f-1$ and for $h_{(2)}^p$, 
$p= 1, \cdots, N_c-N_f-1$. 
Explicit forms for the root generators $E^a$ will not be needed at 
this point.

\par
The potential $V(Q,\bar Q)$ in this basis becomes
\begin{eqnarray}
V
&=&   
\biggl \{ {\rm \, tr} Q_{(1)}^{-2} +{\rm \, tr} \bar Q_{(\bar 1)} ^{-2} +
     {\rm \, tr}  \bar Q_{(\bar 2)} \bar Q_{(\bar 2)} ^\dagger 
\bar Q_{(\bar 1)} ^{-1}
         U^\dagger
         Q_{(1)} ^{-2} U \bar Q_{(\bar 1)} ^{-1}  \biggr \} 
         \biggl [ \prod _{i} 
         Q_{i} \bar Q_{i} \biggr ] ^{-2/(N_c-N_f)} 
\nonumber \\
& &  \quad  
     +  \frac{g^2}{2} \biggl \{ ({\rm \, tr} Q_{(1)} ^2 h_{(1)} ^{p})^2 
       +  ({\rm \, tr} \bar Q_{(\bar 1)} ^2 h_{(1)} ^{p})^2
       - 2 ( {\rm \, tr} Q_{(1)} ^{2} h_{(1)} ^{p}) ({\rm \, tr} U 
\bar Q_{(\bar 1)}^{2}
           U^{\dagger} h_{(1)} ^{p})  
\nonumber \\
& &      \qquad \qquad \quad
       +   ({\rm \, tr} \bar Q_{(\bar 2)}^\dagger \bar Q_{(\bar 2)}
 h_{(2)} ^{q})^{2}
       +  \bigl [ C_{1} {\rm \, tr} (Q_{(1)}^{2}  
-  \bar Q_{(\bar 1)} ^{2})
         + C_{2} {\rm \, tr} \bar Q_{(\bar 2)} ^{\dagger} 
\bar Q_{(\bar 2)}  \bigr ] ^{2}
\nonumber \\   
&&       \qquad  \qquad \quad
          + \bigl [{\rm \, tr} (\bar Q _{(\bar 1)} 
~~\bar Q_{(\bar 2)} ) ^{\dagger} 
            (\bar Q _{(\bar 1)} ~~\bar Q_{(\bar 2)} )  E^{a} \bigr]^{2}
       \biggr \} \nonumber
\\ 
& &   \quad 
     -  m_{Q} ^2 {\rm \, tr} Q_{(1)} ^{2} + m_{\bar Q} ^2 {\rm \, tr} 
\bar Q_{(\bar 1)} ^{2}
       + m_{\bar Q} ^2 {\rm \, tr} \bar Q_{(\bar 2)} 
\bar Q_{(\bar 2)} ^{\dagger}
\label{eq:thirteen}
\end{eqnarray}
We wish to minimize $V$ as a function of the variables $Q_{i}$,
$\bar Q_{i}$, (i.e. the entries of $Q_{(1)}$ and $\bar Q_{(\bar 1)}$ 
respectively), $\bar Q_{(\bar 2)}$ and the $SU(N_{f})$ matrix $U$. 
To do this, we shall show the following simple results.  
\begin{enumerate}
\item 
{}For fixed $Q_{i}$, $\bar Q_{i}$ and $U$, the minimum of $V$
in $\bar Q_{(\bar 2)}$ is at $\bar Q_{(\bar 2)}=0$;
\item 
{}For fixed $Q_{i}$, $\bar Q_{i}$ and $\bar Q_{(\bar 2)}=0$, the minimum
as a function of $U$ is when $U\bar Q_{(\bar 1)} U^\dagger$ is diagonal, 
i.e. a permutation of the entries $\bar Q_{i}$ of $\bar Q_{(\bar 1)}$;
\item 
Under the conditions of 2., and fixed 
$Q_{0} ^{2} = \sum _{i} Q_{i}^{2}/N_{f}$ and $\bar Q_{0}^{2} = \sum _{i}
\bar Q_{i}^{2}/N_{f}$, the minimum of the potential is at  
$Q_{i} = Q_{0}$ and $\bar Q_{i} = \bar Q_{0}$, for all
$i=1,\cdots, N_{f}$. (We have assumed, without loss of generality, that
$Q_{0}\geq 0$, $\bar Q_{0}\geq 0$.)
\item 
The remaining potential is of the same form as the one for $N_{f}=1$, 
and always admits a non-trivial solution with $Q_{0} \not=0$ and
$\bar Q _{0} \not=0$.
\end{enumerate}
 
To show 1., we notice that every term separately involving 
$\bar Q_{(\bar 2)}$ in $V$,
assumes its absolute minimum at $\bar Q_{(\bar 2)}=0$. The only term for 
which this is not a priori obvious is 
$[ C_{1} {\rm \, tr} (Q_{(1)}^{2}  -  \bar Q_{(\bar 1)} ^{2})
+C_{2}{\rm \, tr} \bar Q_{(\bar 2)}^{\dagger} \bar Q_{(\bar 2)}]^{2} $. 
We make use of the fact that with the sign-assignment of the mass terms,
$m_{Q}^2>0$ and $m_{\bar Q}^{2} >0$, we must have 
${\rm \, tr} Q_{(1)}^{2} > {\rm \, tr} \bar Q_{(\bar 1)}^{2}$. 
Since $C_{1}>0$ and $C_{2}>0$, it follows that the above term also 
achieves its absolute minimum when $\bar Q_{(\bar 2)}=0$.

To show 2., we set $\bar Q_{(\bar 2)}=0$, and notice that only a single 
term depends upon the unitary matrix $U$, given by 
$({\rm \, tr}Q_{(1)}^{2} h_{(1)}^{p}) ({\rm \, tr} U\bar Q_{(\bar 1)}^{2}
 U^{\dagger} h_{(1)} ^{p})$. 
Varying with respect to $U$ to find the extrema, we have
\begin{equation}
[h_{(1)} ^p ~, ~ U \bar Q_{(\bar 1)} ^{2} U^\dagger ]~ 
{\rm \, tr} Q_{(1)} ^{2} h_{(1)} ^{p}=0
\label{eq:fourteen}
\end{equation}
Assuming that $Q_{(1)}$ takes on generic values, this equation 
requires that $U\bar Q_{(\bar 1)}^{2}U^\dagger$ should commute with 
all $h_{(1)}^{p}$ matrices, i.e. with all diagonal matrices. 
This is possible only when $U\bar Q_{(\bar 1)}^{2}U^\dagger$ itself is 
diagonal. But then the new matrix will be given
by the matrix $\bar Q_{(\bar 1)}$ in which the entries have been 
interchanged by a
permutation given by $U$. We shall denote this permutation by 
$\sigma _U$ or
just $\sigma$ for short, and its action on $\bar Q_{i}$ is then 
conveniently represented by $\bar Q_{i} \to \bar Q_{\sigma (i)}$. 
To find the absolute minimum amongst the $(N_{f})!$ permutations, 
we make use of the following inner product formulas for the matrices 
$h$, and any two diagonal matrices $A$ and $B$, 
with entries $A_{i}$ and $B_{i}$ respectively :
\begin{equation}
\sum_{p=1}^{N-1} {\rm \, tr} (A h^p) {\rm \, tr} (B h^p) 
= \frac{1}{2N} \sum _{i<j}
(A_{i}-A_{j})(B_{i}-B_{j})
\label{eq:fifteen}
\end{equation}
As a result, the value of the term involving $U$ is re-expressed as
\begin{equation}
-g^2 ( {\rm \, tr} Q_{(1)}^{2} h_{(1)}^{p}) 
({\rm \, tr} U\bar Q_{(\bar 1)}^{2} U^{\dagger}
h_{(1)}^{p}) 
= -\frac{g^2}{2N_{f}} \sum _{i<j} (Q_{i}^{2} - Q_{j}^{2})
(\bar Q_{\sigma (i)} ^{2} - \bar Q_{\sigma (j)} ^{2})
\label{eq:sixteen}
\end{equation}
It is easy to show that the absolute minimum of this term occurs when 
all terms in the sum are separately positive. This happens when the 
orderings of $Q_{i}$ and $\bar Q_{i}$ are correlated as follows :
\begin{eqnarray}
& 
Q_{i_{1}}^{2} \geq Q_{i_{2}}^{2} \geq Q_{i_{3}} ^{2}\geq \cdots \geq
Q_{i_{N_{f}}}^{2}
\nonumber \\ 
& 
\bar Q_{\sigma (i_{1})}^{2} \geq \bar Q_{\sigma (i_{2})}^{2} 
\geq \bar Q_{\sigma (i_3)}^{2} \geq \cdots 
\geq \bar Q_{\sigma (i_{N_{f}})}^{2}
\label{eq:seventeen}
\end{eqnarray}
This completes the argument for point 2.

To show 3., we use (\ref{eq:fifteen}) to express all inner products 
of traces; then, under the conditions of 1. and 2. we have
\begin{eqnarray}
V
&\!\!\!=&\!\!\!  
\sum _i (Q_{i}^{-2} + \bar Q_{i}^{-2} )  \biggl [
\prod _j Q_{j} \bar Q_{j} \biggr ] ^{-2/(N_c - N_f)} 
\nonumber \\
&\!\!\! &\!\!\! 
+ \frac{g^2}{4N_{f}} \sum _{i<j} \biggl \{
    (Q_{i}^{2} - Q_{j}^{2})^{2} + (\bar Q_{i}^{2} - \bar Q_{j}^{2})^2
    -2 (Q_{i}^{2} - Q_{j}^{2})(\bar Q_{\sigma (i)}^{2} - \bar Q_{\sigma
       (j)}^{2})
    \biggr \} 
\nonumber \\
&\!\!\! &\!\!\!
+ g^{2} \frac{N_{c}-N_{f}}{4N_{c}N_{f}} \biggl \{ \sum_{i} Q_{i}^{2} -
 \sum _{i}\bar Q_{i}^{2}\biggr \} ^{2} - m_{Q}^{2} \sum _{i} Q_{i}^{2} +
m_{\bar Q}^{2}
    \sum _{i} \bar Q_{i}^{2}
\label{eq:eighteen}
\end{eqnarray}
Keeping $Q_{0} ^{2} =  \sum _{i} Q_{i}^{2}/N_{f}$,
and $\bar Q_{0}^{2} = \sum _{i} \bar Q_{i}^{2}/N_{f}$ fixed, while 
varying $Q_{i}$ and $\bar Q_{i}$ the terms on the third 
line in eq.(\ref{eq:eighteen}) are fixed under this variation. 
It is easy to see that the non-perturbative term on
the first line of eq.(\ref{eq:eighteen}) assumes its absolute minimum 
when $Q_{i} = Q_{0}$ and $\bar Q_{i} = \bar Q_{0}$ for all 
$i=1,\cdots ,N_{f}$. (Recall that we arranged
all $Q_{i}$ and $\bar Q_{i}$ to be positive.) But, for these values, 
the terms on the second line automatically assume their absolute 
minimum value~: 0. Thus, the configuration where $Q_{i} = Q_{0}$ and
$\bar Q_{i} = \bar Q_{0}$ for all $i=1,\cdots ,N_{f}$, gives the absolute
minimum for the entire potential $V$ subject to the constraint 
$Q_{0} ^{2} =  \sum _{i} Q_{i}^{2}/N_{f}$, and $\bar Q_{0}^{2} 
= \sum _{i} \bar Q_{i}^{2}/N_{f}$, which proves point 3.

To show 4., we substitute the values $Q_{i} = Q_{0}$ and
$\bar Q_{i} = \bar Q_{0}$ for all $i=1,\cdots ,N_{f}$ into the 
expression for $V$ and obtain
\begin{equation}
V/ N_{f}
=
 (Q_{0} ^{2} + \bar Q_{0} ^{2} ) 
\bigl [ Q_{0} \bar Q_{0} \bigr ] ^{-2\gamma }
+ \frac{g^2}{4\gamma} (Q_{0}^{2} - \bar Q_{0}^{2})^{2}
 - m_{Q}^{2} Q_{0}^{2} + m_{\bar Q}^{2} \bar Q_{0}^{2}
\label{eq:nineteen}
\end{equation}
We have introduced a constant that contains all the remaining 
dependence on $N_c$ and $N_f$~:
\begin{equation}
\gamma = \frac{N_{c}}{N_{c}-N_{f}} ~>~1
\label{eq:twenty}
\end{equation}
Clearly, the reduced potential assumes the same form as for a single 
flavor, but with modified value of $\gamma$. 
The minimum conditions are
\begin{eqnarray}
0 &=&
(\gamma -1) Q_{0} ^{-2\gamma}   \bar Q_{0}^{-2\gamma}
  +\gamma     Q_{0} ^{-2-2\gamma} \bar Q_{0}^{2-2\gamma} 
  -\frac{g^2}{2\gamma} (Q_{0}^{2}-\bar Q_{0}^{2}) +m_{Q}^{2} 
\nonumber \\
0 &=&
(\gamma -1) Q_{0} ^{-2\gamma}   \bar Q_{0}^{-2\gamma}
  +\gamma     Q_{0} ^{2-2\gamma} \bar Q_{0}^{-2-2\gamma} 
  +\frac{g^2}{2\gamma} (Q_{0}^{2}-\bar Q_{0}^{2}) -m_{\bar Q}^{2} 
\label{eq:twentyone}
\end{eqnarray}
It is very convenient to introduce new variables $v$ and $x$ with
$0\leq x \leq 1$, in terms of which we have 
\begin{equation}
Q_{0} ^{2}= v^{2}(1+x) \qquad \qquad \bar Q_{0} ^{2} = v ^{2} (1-x)\; ,
\label{eq:twentytwo}
\end{equation}
The equations for $Q_{0}$ and $\bar Q_{0}$ are equivalent to requiring
$v^2=v_{1}^2 (x) = v_{2}^2(x)$, where the functions $v_1^2(x)$ and 
$v_2^2(x)$ are given by
\begin{eqnarray}
v^{2} _{1}(x) 
&=&
(m_{\bar Q}^{2} - m_{Q}^{2})^{-1/2\gamma} (1-x^{2})^{-1/2-1/2\gamma}
  \bigl \{ 4\gamma -2 +2x^{2}\bigr \} ^{1/2\gamma}
\nonumber \\
v^{2}_{2}(x)
&=&
\frac{\gamma}{g^{2}} \biggl \{ \frac{m_{Q}^{2} + m_{\bar Q}^{2}}{2x} -
  \frac{\gamma (m_{\bar Q}^{2} - m_{Q}^{2})}{ 2\gamma-1 +x^2} \biggr \}
\label{eq:twentythree}
\end{eqnarray}
As $x$ ranges from 0 to 1, the function $v_1^2(x)$ is monotonically 
increasing to $\infty$ as $x$ approaches 1, while $v_2^2(x)$ is 
monotonically decreasing from $\infty$ as $x$ approaches $0$. 
{}From this behavior, it is clear that (\ref{eq:twentythree})
admits a unique solution for any arrangement of $g^2>0$,
$\gamma >1$ and masses $m_{\bar Q}^{2} - m_{Q}^{2}>0$.
(Henceforth, we find it convenient to express physical parameters, 
such as masses of various particles, in terms of the original 
dimensionless parameters $\gamma$ and $g$, and the quantities $v$ 
and $x$ determined by the implicit equations of (2.23).)

It is easy to find an approximate solution for the limit where both 
$Q_{0}$ and $\bar Q_{0}$ are large. 
This limit corresponds to $v^2$ large, and according to
(\ref{eq:twentythree}) this means $x$ small : 
\begin{eqnarray}
v^{2}
&\sim & 
  (4\gamma -2)^{1/2\gamma} ~ (m_{\bar Q}^2 - m_{Q}^{2}) ^{-1/2\gamma}  
\nonumber \\
x&\sim &
 \frac{\gamma }{2g^2} 
(4\gamma -2)^{-1/2\gamma} ~ (m_{\bar Q}^{2} + m_{Q}^{2})
  (m_{\bar Q}^2 - m_{Q}^{2}) ^{1/2\gamma} 
\label{eq:twentyfour}
\end{eqnarray}
This situation will occur if the masses $m_{Q}$ and $m_{\bar Q}$ are 
small compared with the non-perturbative scale $\Lambda$, or when 
these masses are large, but only their difference is small compared 
to $\Lambda$.

If we let the mass squared for quarks to increase from negative values 
$-m_Q^2$ to positive values, we find that our solutions smoothly connect 
to the positive  mass squared case in ref.\cite{AhSoPe}. 
Their analysis was done using meson variables appropriate for the 
confining phase, whereas ours uses the colored fundamental 
quark variables appropriate for the Higgs phase. 
This smooth transition gives further evidence for the complementarity of 
confining and Higgs pictures in the case of quarks in the fundamental 
representation \cite{FrSh}. We shall further substantiate this 
complementarity by comparing the mass spectra in Section 2.6. 

%
\vspace{5mm}
\pagebreak[3]
\addtocounter{section}{0}
\addtocounter{subsection}{1}
\setcounter{footnote}{0}
\begin{center}
{\bf \thesubsection. Spectrum and Unbroken Symmetries }
\end{center}
\nopagebreak
\medskip
\nopagebreak
\hspace{2mm}
%
%
It was shown previously that soft supersymmetry breaking squark
mass terms may be introduced, which respect the full $SU(N_{c})\times
SU(N_{f})_{Q}\times SU(N_{f})_{\bar Q}\times U(1)_{B} \times U(1)_{R}$ 
symmetry. Under the assumption that $m_{Q}^{2} <m_{\bar Q}^2$, 
we found that the squark fields acquire non-zero vacuum expectation 
values, 
\begin{equation}
\langle 0 | Q | 0 \rangle = 
\left ( \begin{array}{c} Q_{0} I_{N_f} \\ 0 \\ \end{array} \right ) 
\qquad \qquad \qquad 
\langle 0 | \bar Q | 0 \rangle = 
\left ( \begin{array}{cc} \bar Q_{0}I_{N_f} & 0 \\ \end{array} \right )
\label{eq:twentyfive}
\end{equation}
with $Q_{0}$ and $\bar Q_{0}$ given by eqs.(\ref{eq:twentytwo}) and 
(\ref{eq:twentythree}). In view of these expectation values,
color and flavor symmetries are spontaneously broken to
\begin{equation}
SU( N_{c}-N_f) \times SU(N_{f})_{V} \times U(1)_{B}
\label{eq:twentysix}
\end{equation} 
Here, $SU(N_{f})_{V}$ is the diagonal subgroup of  the $SU(N_{f})$ 
subgroup of color and the flavor group 
$SU(N_{f})_{Q}\times SU(N_{f})_{\bar Q}$. 
The spectrum of the model will transform under linear representations 
of this unbroken group.

In this Section, we shall calculate the masses of the various elementary
constituents, in terms of the breaking parameters evaluated previously. 
We begin by separating the field variables according to irreducible 
representations of $SU( N_{c}-N_f) \times SU(N_{f})_{V}$. 
Different fields can mix only (and will mix in fact) if they have 
the same spin and transform under the same
irreducible representation of $SU( N_{c}-N_f) \times SU(N_{f})_{V}$. 
Thus, the mass matrix can be diagonalized in each respective 
representation subspace. 
The separation of variables is effected as follows
$$
Q  = 
\left ( \begin{array}{c} 
   (Q_{0} +\frac{Q_{(0)}}{\sqrt{N_{f}}})I_{N_{f}} +Q_{(1)} \\ Q_{(2)} \\
        \end{array}
\right )\; , 
\qquad \quad 
\bar Q  = 
\left ( \begin{array}{cc} 
 (\bar Q_{0} +\frac{\bar Q_{(\bar 0)}}{\sqrt{N_{f}}})I_{N_{f}}
 + \bar Q_{(\bar 1)}  & \bar    Q_{(\bar 2)}
\\ 
        \end{array} 
\right )
$$

\begin{equation}
\psi _{Q}  = 
\left ( \begin{array}{c} 
   \frac{\psi_{(0)}}{\sqrt{N_{f}}}I_{N_{f}} +\psi_{(1)} \\ \psi_{(2)} \\
        \end{array}
\right )\; , 
\qquad \quad 
\psi _{\bar Q}  = 
\left ( \begin{array}{cc} 
 \frac{\psi _{(\bar 0)}}{\sqrt{N_{f}}}I_{N_{f}} + \psi _{(\bar 1)} 
&  \psi _{(\bar 2)}
\\ 
        \end{array} 
\right )
\label{eq:twentyseven}
\end{equation}

$$
A_{\mu }   = 
\left ( \begin{array}{cc} 
     A_{\mu (1)} & A_{\mu (2)} \\
     A_{\mu (2)}^{\dagger} & A_{\mu (3)} \\
        \end{array}
\right ) + A_{\mu (0)} H_{(0)} \; , 
\qquad \quad 
\lambda   = 
\left ( \begin{array}{cc} 
     \lambda _{ (1)}           & \lambda _{ (2)} \\
     \lambda _{ (\bar 2)} & \lambda _{ (3)} \\
        \end{array}
\right ) + \lambda _{ (0)} H_{(0)} \; ,  
$$
The fields $Q_{(0)}$, $\bar Q_{(\bar 0)}$, $\psi _{(0)}$, 
$\psi _{(\bar 0)}$, $A_{\mu (0)}$ and $\lambda _{(0)}$ are 
uniquely defined, provided we insist on
the following conditions
\begin{eqnarray}
0&=&
{\rm \, tr}\, Q_{(1)} = {\rm \, tr}\, \bar Q_{(\bar 1)} 
= {\rm \, tr} \psi _{(1)} = {\rm \, tr} \psi _{(\bar 1)} 
\\
\nonumber
0&=&
{\rm \, tr} A_{\mu (1)} = {\rm \, tr} A_{\mu (3)} 
= {\rm \, tr} \lambda _{(1)} = {\rm \, tr} \lambda _{(3)}
\label{eq:twentyeight}
\end{eqnarray}
The transformation properties and multiplicity of these fields are 
summarized in the tables~\ref{table:1}-\ref{table:3} for fields of 
spin 1, spin 1/2 and spin 0 respectively.

\subsubsection{Spin 1 Masses}

Gauge boson mass terms are read off from the $D$-term part of the 
Lagrangian density in eq.(\ref{eq:three}), and are given by
\begin{equation}
-g^{2} A_{\mu}^{p} A^{\mu q} {\rm \, tr} \bigl \{ 
\langle 0 | Q^{\dagger} | 0 \rangle 
t^{p} t^{q} \langle 0 | Q | 0 \rangle 
+ \langle 0 | \bar Q | 0 \rangle 
 t^p t^{q} \langle 0 | {\bar Q}^{\dagger} | 0 \rangle 
\bigr \}
\end{equation}
It is straightforward to read off the masses of the gauge bosons. The
would-be-Goldstone bosons  that are eaten in order to give these vector 
bosons mass are identified by inspecting  the linear coupling of the 
scalar fields to the gauge bosons.  These couplings are given by
\begin{eqnarray}
 ig &\{\!\!\! & \!\!\! \frac1{\sqrt{\gamma}} A_{\mu (0)} \partial ^{\mu} 
      ( Q_{0} Q_{(0)} + \bar Q_0 \bar Q_{(\bar 0)}^*  )
     + {\rm \, tr} A_{\mu (1)}  \partial ^{\mu} 
      ( Q_{0} Q_{(1)} + \bar Q_0 \bar Q_{(\bar 1)}^\dagger  )
\nonumber \\ 
&+\!\!\!&     {\rm \, tr} A_{\mu (2)}  \partial ^{\mu} 
  ( Q_{0} Q_{(2)} + \bar Q_0 \bar Q_{(\bar 2)}^\dagger  ) \} + {\rm c.c.}
\label{eq:thirty}
\end{eqnarray}
The properties of the vector bosons are summarized in the 
table~\ref{table:1}.

\begin{table}[t]
\begin{center}
\begin{tabular}{|c|c|c|c|} \hline 
Spin $1$  & Masses & Multiplicity  & $SU(N_f)\times
SU( N_c-N_f)$
                \\ \hline \hline
$A_{\mu (0)}$ & $gv\sqrt{2/\gamma}$ & 1 & $1 \otimes 1$ 
                \\ \hline
$A_{\mu (1)}$ & $gv\sqrt{2}$ & $N_{f}^2 -1$ & adjoint $\otimes \, 1$
                \\ \hline
$A_{\mu (2)}$ & $gv$ & $2N_{f}(N_{c}-N_f)$ & $N_{f}^{*}\, \otimes \,
 (N_c-N_f)~\oplus~ N_{f}\, \otimes \, (N_c-N_f)^{*}$ 
                \\ \hline
$A_{\mu (3)}$ & 0 & $ (N_{c}-N_f)^{2} -1$ & $1 \, \otimes$ adjoint
                \\ \hline
\end{tabular}
\end{center}
\caption{The spin 1 fields and their masses ($N_f < N_c$)}
\label{table:1}
\end{table}

\subsubsection{Spin 1/2 Masses}

The fermion masses are governed by contributions both from the 
superpotential (contribution $V_1$) and from the mixing of the 
quarks and the gauginos in the $D$-term (contribution $V_2$) in 
eq.(\ref{eq:three}). The first contribution, in
terms of the irreducible fields of eq.(\ref{eq:twentyseven}) is 
given by
\begin{eqnarray}
V_{1} & = &
\frac12 \gamma M \psi _{(0)}\psi _{(0)} +
\frac12 \gamma \bar M \psi _{(\bar 0)}\psi _{(\bar 0)} +
(\gamma -1) \sqrt{M \bar M} \psi _{(0)}\psi _{(\bar 0)} 
\nonumber \\
&& + 
\frac12 M {\rm \, tr} \psi _{(1)}\psi _{(1)} +
\frac12 \bar M {\rm \, tr} \psi _{(\bar 1)}\psi _{(\bar 1)} +
\sqrt{M\bar M} {\rm \, tr} \psi _{(\bar 2)}\psi _{(2)} + {\rm c.c.}
\label{eq:thirtytwo}
\end{eqnarray}
Here, we have introduced the notation $M$ and $\bar M$ for the 
effective masses coming from the non-perturbative potential. They can be
expressed as follows in terms of the vacuum expectation values $Q_{0}$, 
$\bar Q_0$ or in terms of the basic parameters~:
\begin{eqnarray}
M&=&
 Q_{0} ^{-\gamma -1} \bar Q_{0} ^{-\gamma +1} 
= (1-x) (m_{\bar Q} ^2 - m_Q ^{2})^{1/2} (4\gamma -2 +2 x^2) ^{-1/2}
\nonumber \\
\bar M&=&
 Q_{0} ^{-\gamma +1} \bar Q_{0} ^{-\gamma -1}
= (1+x) (m_{\bar Q} ^2 - m_Q ^{2})^{1/2} (4\gamma -2 +2 x^2) ^{-1/2}
\end{eqnarray}
Similarly, the contribution $V_2$ may be decomposed in terms of the
irreducible fields of eq.(\ref{eq:twentyseven}), and we obtain
\begin{eqnarray}
V_2 & =&
-i g/\sqrt{\gamma} Q_{0} \lambda _{(0)} \psi _{(0)} 
+i g/\sqrt{\gamma} \bar Q_{ 0} \lambda _{(0)} \psi _{(\bar 0)} 
\nonumber \\
&&
-i\sqrt{2} g Q_{0} \bigl ({\rm \, tr} \lambda _{(1)} \psi _{(1)} 
                        + {\rm \, tr} \lambda _{(2)} \psi _{(2)} \bigr ) 
\nonumber \\
&&
+i\sqrt{2} g \bar Q_{0} 
\bigl ({\rm \, tr} \lambda _{(1)} \psi _{(\bar 1)} 
              + {\rm \, tr} \lambda _{(\bar 2)} \psi _{(\bar 2)} \bigr )
+ {\rm c.c.}
\label{eq:thirtythree}
\end{eqnarray}
{}From these mass terms, it is clear that the following groups of 
fields mix amongst each other, but do not mix from one group to 
the next. Singlets under $SU(N_f)$~: 
($\psi _{(0)},~\psi_{(\bar 0)},~ \lambda _{(0)}$); adjoint under
$SU(N_f)$ : ($\psi _{(1)},~\psi_{(\bar 1)},~ \lambda _{(1)}$); 
fundamental under
$SU(N_f)$ : ($\psi _{(2)},~\psi_{(\bar 2)},~ \lambda _{(2)},~\lambda
_{(\bar 2)}$). Thus, the mass matrix may be diagonalized in each set
separately. 

In the set ($\psi _{(0)},~\psi_{(\bar 0)},~ \lambda _{(0)}$), the
masses are governed by the cubic equation
\begin{eqnarray}
&0=&
{\cal M}^3 - {\cal M}^2 \gamma (M+\bar M) +{\cal M} 
\{(2\gamma-1)M\bar M + 2\frac{g^2}{\gamma}v^2\} 
\nonumber \\
&&
 -g^2(\bar Q_{0} \sqrt {M} +Q_{0} \sqrt{\bar
M})^2+2\frac{g^2}{\gamma}Q_{0}\bar Q_{0} \sqrt {M\bar M} 
\label{eq:thirthyfour}
\end{eqnarray}
It is instructive to evaluate the solutions in the limit where 
$M,~\bar M \ll gv$, and we find the masses
\begin{eqnarray}
{\cal M}_{(0)}^{0}   
&=& \frac{\gamma}{2v^2} (\bar Q_{0} \sqrt {M} +Q_{0} \sqrt{\bar
M})^2 -\frac{1}{v^2}Q_{0}\bar Q_{0}\sqrt {M\bar M} +{\cal O}(M^2) 
\nonumber \\
{\cal M}_{(0)}^{\pm} 
&=& \pm \frac{i\sqrt{2}}{\sqrt{\gamma}}gv +{\gamma \over 2}  
(M+\bar M) -
\frac12 {\cal M}_{(0)}^{0} +{\cal O}(M^2) 
\label{eq:thirtyfive}
\end{eqnarray}
The field $\psi _{(0)} ^{0}$, with mass ${\cal M}_{(0)}^{0}$, that is 
light in this limit, is given approximately by
\begin{equation}
\psi _{(0)} ^{0} = \frac{1}{\sqrt{2} v} (\bar Q_{0} \psi _{(0)} + Q_0 
\psi_{(\bar 0)})
\end{equation}
while the other two heavy fields correspond to the orthogonal linear
combinations.

In the set ($\psi _{(1)},~\psi_{(\bar 1)},~ \lambda _{(1)}$), 
the masses of the three irreducible representations again satisfy a cubic
equation,  given by
\begin{equation}
0={\cal M} ^3 -{\cal M} ^2 (M+\bar M) + {\cal M} M\bar M +4 g^2 v^2 
{\cal M} 
-2 g^2(\bar Q_0 ^2 M + Q_0^2\bar M)
\end{equation}
In the limit where $M,~\bar M \ll gv$, we find the masses
\begin{eqnarray}
{\cal M}_{(1)}^{0}   
&=&
 \frac{1}{2v^2} (\bar Q_{0} ^2 M +Q_{0}^2 \bar M)+{\cal O}(M^2) 
\nonumber \\
{\cal M}_{(1)}^{\pm} 
&=&
 \pm 2igv+ \frac12 (M+\bar M)- \frac{1}{2} {\cal M}_{(1)}^{0} +{\cal O}
(M^2), 
\quad 
\label{eq:thirtysix}
\end{eqnarray}
The field $\psi _{(1)} ^{0}$ of mass ${\cal M}_{(1)}^{0}$, that is light 
in this limit is given approximately by
\begin{equation}
\psi _{(1)} ^{0} 
= \frac{1}{\sqrt{2} v} ( \bar Q_{0} \psi _{(1)} + Q_{0} \psi
_{(\bar 1)})
\end{equation}

In the set ($\psi _{(2)},~\psi_{(\bar 2)},~ \lambda _{(2)},
~\lambda _{(\bar 2)}$), the masses of the four irreducible 
representations  satisfy a quartic equation which is even, given by
\begin{equation}
0= {\cal M} ^4 +{\cal M} ^2 ( 4g^2 v^2 - M\bar M) +4 g^4 Q_{0}^{2} \bar
Q_{0}^{2}
\end{equation}
In the limit where $M,~\bar M \ll gv$, we find 
\begin{eqnarray}
\pm {\cal M}_{(2)} 
&=&
 \pm i\sqrt{2} gQ_{0}^{2} +{\cal O}(M^2), 
\\
\quad 
\pm \bar {\cal M}_{(2)} 
&=&
 \pm i\sqrt{2} g\bar Q_{0}^{2} +{\cal O}(M^2) 
\nonumber 
\label{eq:thirtyseven}
\end{eqnarray}
{}From the mass eigenvalue equations for fermions, it is evident that 
${\cal M}=0$ eigenvalues can never occur so that all fermions are always 
massive.

A summary of all spin 1/2 fields and masses is given in the 
table~\ref{table:2}.

\begin{table}[t]
\begin{center}
\begin{tabular}{|c|c|c|c|} \hline 
Spin $1/2$  & Masses & Multiplicity  & $SU(N_f)
\times SU( N_c-N_f)$ 
  \\ \hline \hline
$\psi_{(0)},~\psi_{(\bar 0)},~\lambda_{(0)}$ 
& ${\cal M}_{(0)} ^{\pm},~ {\cal M}_{(0)} ^{0}$
   & $3 $ & $1\otimes 1$ 
  \\ \hline
$\psi_{(1)},~\psi_{(\bar 1)},~\lambda_{(1)}$ 
& ${\cal M}_{(1)} ^{\pm},~ {\cal M}_{(1)}^{0}$
   & $3  (N_{f}^2-1)$ &  adjoint  $\otimes \, 1$
  \\ \hline
$\psi_{(2)},~\lambda_{(2)}$,  
& $\pm {\cal M}_{(2)} $
   & $2 N_f (N_c-N_f)$ & $2N_f ^{*}\, \otimes\, (N_c-N_f)~  $   
  \\ 
$\psi_{(\bar 2)},~\lambda_{(\bar 2)}$ 
& $ \pm \bar {\cal M}_{(2)}$ & $2 N_f (N_c-N_f)$ &$ 2N_{f}\, \otimes \, 
(N_c-N_f)^{*}$  
  \\ \hline
\end{tabular}
\end{center}
\caption{The spin 1/2 fields and their masses ($N_f < N_c$)}
\label{table:2}
\end{table}

\subsubsection{Spin 0 Masses}

{}Finally, the masses of spin 0 bosons are again found by decoupling the
equations according to the fields of eq.(\ref{eq:twentyseven}). 
There are three
groups of irreducible representations : the singlets
under $SU(N_f)$ : $(Q_{(0)},~\bar Q_{(\bar 0)})$, 
the adjoints under $SU(N_f)$~: $(Q_{(1)},~\bar Q _{(\bar 1)})$ and the
fundamental under $SU(N_f)$ :
$(Q_{(2)},~\bar Q_{(\bar 2)})$. The dynamics of the scalar fields 
is governed
by the potential $V$ of eq.(\ref{eq:eight}), 
and we shall now analyze the mass
spectrum for each of the irreducible groups in turn.

The dynamics for the group $(Q_{(0)},~\bar Q_{(\bar 0)})$ may be split as
follows. First, the phases of $Q_{(0)}$ and $\bar Q_{(\bar 0)}$ are 
massless. 
One of them is eaten by the massive gauge field $A_{\mu (0)}$, while the 
other is a true Goldstone boson, and has zero mass. 
Second, the masses of the norms of these fields is given by 
the reduced potential of (\ref{eq:nineteen}), and thus characterized by 
a quadratic equation
\begin{eqnarray}
0&=&
\bigl ({\cal M} ^2 -2\gamma (\gamma -1) M\bar M -2\gamma (\gamma +1) 
M^2 - 
\frac{g^2}{\gamma} Q_{0} ^2 \bigr )\times
\nonumber \\
&&
\bigl ({\cal M} ^2 -2\gamma (\gamma -1) M\bar M -2\gamma (\gamma +1) 
\bar M^2 - 
\frac{g^2}{\gamma} \bar Q_{0} ^2  \bigr ) 
\nonumber \\
&&
\qquad 
-\bigl ( 4 \gamma (\gamma -1) \frac{v^2}{Q_0\bar Q_{0}} M\bar M
-\frac{g^2}{\gamma} Q_0\bar Q_0 \bigr ) ^2
\end{eqnarray}
In the limit where $M,~\bar M \ll gv$, the solutions reduce to
\begin{eqnarray}
{\cal M} _{(0)}^{0}  
&=&
\sqrt{ \gamma (2\gamma -1)} (M+\bar M) +{\cal O}(M^3) \nonumber \\
{\cal M} _{(0)}  
&=&
 \frac{\sqrt{2}}{\sqrt{\gamma}} gv + {\gamma ^{3/2} (M+\bar M)^2 \over 2
\sqrt{2} gv}  +{\cal O}(M^4)
\label{scalarsinglet}
\end{eqnarray}

The dynamics of the second group $(Q_{(1)},~\bar Q _{(\bar 1)})$ can be 
split further into three different components. From the couplings to the 
gauge fields, given in (\ref{eq:thirty}), it appears that the following 
linear combination 
\begin{equation}
\frac{1}{2v} \bigl \{ Q_{0} (Q_{(1)} -  Q_{(1)} ^{\dagger}) + \bar Q_{0} 
(\bar Q_{(\bar 1)} - \bar Q_{(\bar 1)} ^{\dagger}) \bigr \}
\end{equation}
corresponds to the would-be-Goldstone bosons that are eaten by the gauge 
field $A_{\mu(1)}$. Analysis of the spontaneous symmetry breaking 
mechanism in the potential further reveals that the following linear 
combination
\begin{equation}
\frac{1}{2v} \bigl \{ -\bar Q_{0} (Q_{(1)} -  Q_{(1)} ^{\dagger}) + Q_{0}
(\bar Q_{(\bar 1)} - \bar Q_{(\bar 1)} ^{\dagger}) \bigr \}
\end{equation}
corresponds to the true massless Goldstone bosons, transforming under the
adjoint representation of $SU(N_{f})$. The remaining combinations of
$Q_{(1)}+Q_{(1)}^{\dagger}$ and $\bar Q_{(1)} +\bar Q_{(1)} ^{\dagger}$ 
have masses determined by a quadratic equation, given as follows
\begin{eqnarray}
0&=& 
\bigl ( {\cal M} ^2 -2(\gamma -1) M \bar M - 2(\gamma +1) M^2 
- g^2 Q_0 ^2 \bigr)
\times
\nonumber \\ 
&&
\times \bigl ( {\cal M} ^2 -2(\gamma -1) M \bar M 
- 2(\gamma +1) \bar M^2 - 
 g^2
\bar Q_0 ^2 \bigr )
-g^4 Q_0^2 \bar Q_{0}^2
\end{eqnarray}
In the limit where $M,~\bar M \ll gv$, the solutions reduce to
\begin{eqnarray}
{\cal M} _{(1)}^+  
&=&
 \sqrt{2} gv + \frac{\gamma (M+\bar M)^2}{ 2 \sqrt{2} gv} +{\cal O}(M^4)
\nonumber \\
{\cal M} _{(1)}^-  
&=&
\sqrt{\gamma} (M+\bar M) + {\cal O}(M^3)
\label{scalaradjoint}
\end{eqnarray}

The dynamics of the third group $(Q_{(2)},~\bar Q _{(\bar 2)})$ further 
splits into two components. From the couplings to the gauge field in
(\ref{eq:thirty}), it appears that the linear combination (and its 
hermitian conjugate) defined as follows
\begin{equation}
\frac{1}{\sqrt{2}v} (Q_0 Q_{(2)} + \bar Q_{0} \bar Q_{(\bar 2)}^\dagger )
\end{equation}
corresponds to the would-be-Goldstone bosons eaten by the gauge field
$A_{\mu(2)}$ (and its hermitian conjugate). The remaining linear 
combination
\begin{equation}
\frac{1}{\sqrt{2}v} (-\bar Q_0 Q_{(2)} 
+  Q_{0} \bar Q_{(\bar 2)}^\dagger )
\end{equation}
has mass given by the following exact formula
\begin{equation}
{\cal M} _{(2)} = \sqrt{g^2 v^2 + \gamma (M+\bar M)^2}
\end{equation}

The results on spin 0 boson masses are summarized in the 
table~\ref{table:3}.
The masses of the would-be-Goldstone bosons are referred to by the 
gauge fields into which they have combined, so the corresponding masses 
can be found in the table~\ref{table:1}.

\begin{table}[t]
\begin{center}
\begin{tabular}{|c|c|c|c|} \hline 
Spin $0$  & Masses & Multiplicity  & $SU(N_f)\times
SU( N_c-N_f)$
                \\ \hline \hline
$Q_{(0)},~\bar Q_{(0)}$  &  $[A_{\mu (0)}]$,~0,~ ${\cal M}_{(0)}^{\pm}$ 
&  4  &
   $1\otimes 1$ 
                \\ \hline
$Q_{(1)},~\bar Q_{(\bar 1)}$  
&  $[A_{\mu (1)}]$,~0,~${\cal M}_{(1)}^{\pm}$  &
   $4(N_{f}^2-1)$   &  adjoint $\otimes \, 1$
                \\ \hline
$Q_{(2)},~\bar Q_{(\bar 2)}$  &  $[A_{\mu (2)}]$,~${\cal M}_{(2)}$  &
   $4N_{f}( N_{c}-N_f)$  &  $N_{f}^{*} \, \otimes \, 
( N_{c}-N_f)~ \oplus $ c.c.
                \\ \hline
\end{tabular}
\end{center}
\caption{The spin 0 fields and their masses ($N_f < N_c$)}
\label{table:3}
\end{table}

\vspace{5mm}
\pagebreak[3]
\addtocounter{section}{0}
\addtocounter{subsection}{1}
\setcounter{footnote}{0}
\begin{center}
{\bf \thesubsection. Complementarity between confining and Higgs phase }
\end{center}
\nopagebreak
\medskip
\nopagebreak
\hspace{2mm}
Let us compare our results with those in ref.~\cite{AhSoPe}. There, 
the meson fields $T$ are used as fundamental variables in the low energy 
effective theory. 
This choice of variables is most appropriate to describe the 
physics in the confining phase. 
On the other hand, in the present paper, we have used the fundamental
quark fields $Q$ and $\bar Q$ instead, which are more appropriate for 
the description of the physics of the Higgs phase. 
In nonsupersymmetric gauge theories, with scalar fields in the 
fundamental representation, it was argued that the Higgs and confining
phases are smoothly connected to one another over at least some region 
of coupling constants \cite{FrSh}. 
Therefore, one should be able to describe the same physics in terms of
variables appropriate for either the confining phase or those for 
Higgs phase. 

Thus, we expect that our results should be reproduced using the 
field variables most suitable for the confining phase, as used in
ref.~\cite{AhSoPe}. 
Naturally, this comparison will only hold in the limit of small
supersymmetry breaking mass terms, 
where the approximations of ref.~\cite{AhSoPe} are valid, 
and in the limit of the low mass part of the spectrum. 

If we use the meson fields $T$ which are appropriate to describe the 
physics of confining phase, we find the following mass spectrum
\footnote{
Here, we have made numerical corrections compared 
to the original results in ref~\cite{AhSoPe}. 
However, the conclusions reached in this subsection are not dependent 
upon these corrections in an essential way.}
in the above limit. 
\begin{enumerate}
\item The spin $1/2$ masses for singlet and adjoint are 
\begin{equation}
m_{\rm f, singlet} = \frac{\gamma}{k} t^{-(1+\gamma)}, \qquad 
m_{\rm f, adjoint}  =  \frac{1}{k} t^{-(1+\gamma)}.
\end{equation}
\item The spin 0 masses for singlet and adjoint are 
\begin{equation}
m^2_{\rm s, singlet}  
=  \frac{\gamma (\gamma+1)}{k^2} t^{-2(1+\gamma)}, \qquad
m^2_{\rm s, adjoint}  =   \frac{\gamma+1}{ k^2} t^{-2(1+\gamma)}.
\end{equation}
where the vacuum expectation value of $T$ is denoted as $t$. 
Here we have employed the Ansatz of the K\"ahler potential 
for $T$ following ref.~\cite{AhSoPe} 
\begin{equation}
K[\hat T] = k {\rm \, tr} \hat T^\dagger \hat T.
\label{kaehlerpotfort}
\end{equation}
\end{enumerate}

If we use the fundamental quark fields $Q, \bar Q$ on the other hand, 
which are appropriate to describe the physics of the Higgs phase, 
we obtain the following low energy mass spectrum.\footnote{
In the limit of small supersymmetry breaking parameters, we have 
$M, \bar M \ll gv$.} 
\begin{enumerate}
\item The spin $1/2$ masses for singlet and adjoint are obtained 
from the first of eqs.(\ref{eq:thirtyfive}) and (\ref{eq:thirtysix})
\begin{equation}
m_{\rm f, singlet}= (2 \gamma-1) M, \qquad
m_{\rm f, adjoint}=              M. 
\end{equation}
\item The spin 0 masses for singlet and adjoint are  obtained 
from the first of eq.(\ref{scalarsinglet}) and second of 
eq.(\ref{scalaradjoint})
\begin{equation}
m^2_{\rm s, singlet} = 4 \gamma (2\gamma -1) M^2, \qquad
m^2_{\rm s, adjoint} = 4 \gamma  M^2.
\end{equation}
\end{enumerate}

We see that the qualitative features of the two descriptions are 
the same, confirming the validity of complementarity between the Higgs 
and confining phases at a qualitative level.
However, we find a discrepancy between the results in the two 
descriptions upon closer inspection.
The mass ratios of the singlet and adjoint particles 
in the two description are different, both for 
spin $1/2$ as well as spin $0$ particles. 
Although the physical values of the various parameters in the 
two descriptions cannot be directly related, quantities like the mass 
ratios should be free of any ambiguities.

The resolution of this discrepancy lies in the following simple remarks. 
The K\"ahler potential receives qunatum corrections which cannot be 
controlled by holomorphy. 
Therefore the simple Ansatz (\ref{kaehlerpotfort}) of ref.~\cite{AhSoPe} 
may not be valid. 
Moreover, when spontaneous symmetry breaking occurs, and there is 
a nonvanishing vacuum expectation value for some of the scalar fields, 
naive dimensional counting of operators has to be modified. 
In particular, if we are interested in the approximation with at most 
two derivatives for boson fields and one derivative for fermion fields, 
then we need to keep all contributions to the K\"ahler potential in $T$ 
that only involve the fields $T$ and $T^\dagger$. Thus, the K\"ahler 
potential used in (2.55) and in ref.~\cite{AhSoPe}, is not expected 
to be sufficient to reproduce completely the low energy behavior of
the model, 
or the low mass part of the spectrum.
Instead, we need to include higher order terms in the K\"ahler 
potential as well in order to obtain the quadratic terms correctly,  
\begin{equation}
K[T]= k_1 {\rm \, tr} T^\dagger T + k_2 {\rm \, tr} (T^\dagger T)^2
     +k_3 ({\rm \, tr} T^\dagger T)^2 + \cdots ~.
\end{equation}
It is easy to see that the parameters $k_2, k_3$ which are of order 
$\Lambda^{-2}$, modify the coefficient of the kinetic term as well 
as the masses of spin $1/2$ and $0$ particles, in the confining 
description of ref.~\cite{AhSoPe}. 
In the limit of large vacuum expectation values, we have a nearly 
perturbative situation where our assumption for the K\"ahler 
potential of the quark superfields is reliable. 
In this limit, we can find the corresponding K\"ahler potential 
for the meson field explicitly as 
$K[T]=2 {\rm tr} \left[\left(T^\dagger T\right)^{1/2}\right]$ 
along the D-flat direction. 

In principle, similar corrections to the K\"ahler potential also occur 
even if we use the fundamental fields $\hat Q$, $\hat {\bar Q}$ and 
$\hat V$ which are appropriate to describe the Higgs phase physics. 
However, there is a well-defined limit, the semi-classical limit, 
given by $\hbar \to 0$, where such corrections vanish. 
In this limit, the vacuum is determined by the minimum 
of the classical potential, and the masses are just given 
by the classical fluctuations about this minimum, which is 
exactly what we used here. 
%
\vspace{7mm}
\pagebreak[3]
\addtocounter{section}{1}
\setcounter{equation}{0}
\setcounter{subsection}{0}
\setcounter{footnote}{0}
\begin{center}
{\large {\bf \thesection. Dynamics for $N_f > N_c+1$ }}
\end{center}
\nopagebreak
\medskip
\nopagebreak
\hspace{3mm}
%

In this section, we shall consider the case of $N_f > N_c+1$. 
It is most convenient to use the dual variables instead 
of the original quark and antiquark superfields $\hat Q, \hat{\bar Q}$ 
in this case \cite{Seiberg}. 
The dual description for the gauge group $SU(N_c)$
and $N_f$ flavors of quarks and antiquarks (with $N_f > N_c+1 $) 
has a gauge group $SU(\tilde{N}_c)$ with $N_f$ flavors,
where $\tilde{N}_c = N_f - N_c$. The elementary chiral superfields
in the dual theory are dual quark $\hat q$ 
and meson $\hat T$ superfields,
\begin{equation}
\hat{q}^a{}_i \qquad \hat{\bar{q}}^i{}_a \qquad \hat{T}^i{}_j \qquad 
\quad
a=1, \cdots, \tilde{N}_c; \qquad i,j=1, \cdots N_f,
\end{equation}
which contain scalar fields $q$, $\bar q $, $T$, and left-handed
spinor fields $\psi _q$, $\psi _{\bar q}$ and $\psi _T$ respectively.

The dual theory has color $SU(\tilde N_c)$ gauge invariance, as well as  
the same global 
$G_f=SU(N_f)_Q \times SU(N_f)_{\bar{Q}} \times U(1)_B \times U(1)_R$
symmetry as the original $SU(N_c)$ theory. 
The bosonic and left-handed fermionic components 
transform under these symmetry transformations as
\begin{eqnarray}
{q} \to & U_c ~{q} ~U_Q^T  \qquad \qquad U_Q \in 
& SU(N_f)_Q, \nonumber \\
{\bar{q}}  \to & U_{\bar Q}^* ~{\bar{q}}~ U_c^\dagger  
        \qquad \qquad U_{\bar Q} \in & SU(N_f)_{\bar Q}, \\
{T} \to & U_Q^* ~{T}~ U_{\bar Q}^{T}
        \qquad\qquad U_c \in & SU(\tilde N_c) \; , \nonumber
\end{eqnarray}
with baryon number charges ${N_c \over \tilde N_c}$, 
$-{N_c \over \tilde N_c}$ and 0 respectively, and $R$-charges 
given by
\begin{eqnarray}
&
{q}, ~{\bar q} ~ : ~ N_c/N_f 
\qquad &\qquad
T ~ : ~ 2-2 N_c/N_f 
\nonumber \\
&
\psi_{q}, ~\psi_{\bar q} ~ : ~ N_c/N_f -1
\qquad &\qquad
\psi_T ~ : ~ 1-2 N_c/N_f 
\end{eqnarray}
A further classical axial baryon number $U(1)_{AB}$ symmetry suffers a
$SU(\tilde N_c)$ color anomaly and is absent at the quantum level.
\vspace{5mm}
\pagebreak[3]
\addtocounter{section}{0}
\addtocounter{subsection}{1}
\setcounter{footnote}{0}
\begin{center}
{\bf \thesubsection. The Effective Potential for Scalar Fields}
\end{center}
\nopagebreak
\medskip
\nopagebreak
\hspace{2mm}

Since $\hat q$, $\hat {\bar q}$ and $\hat T$ are effective fields, their 
kinetic terms need not have canonical normalizations; in particular, 
they can receive nonperturbative quantum corrections. Thus, we introduce 
into the (gauged) K\"{a}hler potential for $\hat q$, $\hat {\bar q}$ 
and $\hat T$ normalization parameters $k_q$ and $k_T$ as follows  
\begin{equation}
K[\hat q,\hat {\bar q},\hat T,\hat v]
= k_q {\rm tr} (\hat q^{\dagger} e^{2\tilde
g \hat v} \hat q  + \hat {\bar q} e^{-2\tilde g \hat v} 
\hat {\bar q}^{\dagger})
                 + k_T {\rm tr} \hat T^{\dagger} \hat T.
\label{kaehler}
\end{equation}
Here, we denote by $\hat v$ the $SU(\tilde N_c)$ color gauge superfield, 
and by $\tilde g$ the associated coupling constant.  (Pure gauge terms 
will not be exhibited explicitly.) In principle, these normalization 
parameters are determined by the dynamics of the underlying microscopic 
theory.
{}Furthermore, it has been pointed out in \cite{Seiberg} that a 
superpotential
coupling $q$, $\bar q$ and $T$ should be added as follows 
\begin{equation}
W= \hat{q}^a{}_i \hat{T}^i{}_j \hat{\bar{q}}^j{}_a.
\end{equation}

We add soft supersymmetry breaking terms 
to the Lagrangian for the dual quark and meson supermultiplets.
Besides gaugino masses, the possible soft terms are 
\begin{equation}
{\cal L}_{sb}=
- {\rm tr} (q M_q^2 q^{\dagger} + \bar{q}^{\dagger} M_{\bar{q}}^2 
\bar{q} 
) 
- M^2_T{}^i{}_j{}^k{}_l T^j{}_i T^\dagger{}^l{}_k 
- (A_i{}^j{}_k{}^l \bar{q}^i{}_a q^a{}_j T^k{}_l  + {\rm h.c.}). 
\label{dualsoftterm}
\end{equation}
Here $A$ is a cubic coupling,\footnote{
In general this is a four-index tensor in flavor space, 
but becomes proportional to $\delta_i^l \delta_k^j$ in popular models 
for soft breaking terms using the supergravity \cite{Nilles}.} 
and $M_q^2$, $M_{\bar{q}}^2$ and
$M_T^2$ are mass matrices of $q$, $\bar{q}$ and $T$ respectively.
The $R$-symmetry is broken by the $A$-term in eq.(\ref{dualsoftterm}) 
and would also be broken by gaugino masses. 

{}For simplicity we shall assume that $R$-symmetry is maintained so 
that neither $A$-terms  nor gaugino masses are present in the 
Lagrangian. 
Putting together all $F$ and $D$ terms as well as supersymmetry breaking 
scalar mass terms, the scalar potential is given by 
\begin{eqnarray}
V(q,\bar{q},T) &=& \frac1{k_T} {\rm tr} (q q^{\dagger} \bar{q}^{\dagger}
\bar{q}) + \frac1{k_q} {\rm tr} (q T T^{\dagger} q^{\dagger}
+ \bar{q}^{\dagger}
T^{\dagger} T \bar{q}) 
+\frac{\tilde g^2}2 ({\rm tr} q^{\dagger} \tilde t^a q - {\rm tr} \bar{q}
\tilde t^a 
\bar{q}^{\dagger})^2  \nonumber \\
&& + {\rm tr} (q^{\dagger} M_q^2 q + \bar{q} M_{\bar{q}}^2 
\bar{q}^{\dagger} + T^{\dagger} M_T^2 T). 
\end{eqnarray}
where $\tilde t^a$ denote generators of $SU(\tilde N_c)$.
\vspace{5mm}
\pagebreak[3]
\addtocounter{section}{0}
\addtocounter{subsection}{1}
\setcounter{footnote}{0}
\begin{center}
{\bf \thesubsection. Vacuum Stability }
\end{center}
\nopagebreak
\medskip
\nopagebreak
\hspace{2mm}
%

When the eigenvalues of $M_q^2$, $M_{\bar q}^2$ and $M_T^2$ can take 
generic positive or negative values, the scalar potential may be 
unbounded from below.
A necessary condition for which the potential is bounded from below is 
that $M_T^2$ be a positive definite matrix.
This is because there is no quartic term of $T$.

The D terms vanish when the vacuum expectation values are given by:
\begin{equation}
\langle 0| q |0\rangle = \left(
 \begin{array}{ccccc}
 q_1 &        &                 &   & \\
     & \ddots &                 & 0 & \\
     &        & q_{\tilde{N}_c} &   &
 \end{array}
\right), \qquad
\langle 0| {\bar q}|0 \rangle = \left(
 \begin{array}{ccc}
 \bar{q}_1 &        &                       \\
           & \ddots &                       \\
           &        & \bar{q}_{\tilde{N}_c} \\
           &        &                       \\
           &   0    &   
 \end{array}
\right), 
\end{equation}
with the combinations $|q_i|^2 - |\bar{q}_i|^2$ independent of $i$.

If we set the squark masses to be zero,
the space where $|q_i|^2$ is independent of $i$ and 
$\bar{q}=0$ is a subspace of the moduli space of vacua. 
If we insist on flavor symmetric mass squared matrix and on having a 
negative eigenvalue, we are forced to have a potential unbounded 
from below. In fact, in the next subsection, we shall establish more 
generally
that to have a potential bounded from below, we must have 
\begin{equation}
m_1^2 + \cdots + m_{\tilde{N}_c}^2 \geq 0,
\end{equation} 
where $m_i^2$ are eigenvalues of the matrix $M_q^2$ or $M_{\bar q}^2$,
and they are set to be $m_1^2 \leq m_2^2 \leq \cdots \leq m_{N_f}^2$.

Therefore we consider the simplest stable situation, where 
the $n$ eigenvalues of $M^2_q$ is negative and same, 
while all the others are positive or
zero. The $n$ should be smaller than $\tilde{N}_c$.
{}For simplicity we shall also assume that the soft supersymmetry 
breaking positive mass squared terms for squarks have a flavor symmetry 
$SU(N_{f}-n)_Q \times SU(N_{f})_{\bar Q}$. 
As a result, the $N_{f}-n$
positive  eigenvalue of $M_q^2$ are all the same, while the $N_f$ 
eigenvalues of  $M_{\bar q}^2$ are the same 
:$M_{\bar q}^2=m_{\bar q}^2 I_{N_{f}}$. 
We also assume 
$M_T^2{}^i{}_j{}^k{}_l=m_T^2 \delta^i_j \delta^k_l$.

\vspace{5mm}
\pagebreak[3]
\addtocounter{section}{0}
\addtocounter{subsection}{1}
\setcounter{footnote}{0}
\begin{center}
{\bf \thesubsection. Determination of VEV }
\end{center}
\nopagebreak
\medskip
\nopagebreak
\hspace{2mm}

We begin by making a convenient choice of parameterization for the 
variables $q$ and $\bar{q}$.
We can always represent the vacuum expectation values of 
$q$, $\bar{q}$ and $T$ as follows;
\begin{eqnarray}
\langle 0| q |0 \rangle       &=& U_{c(1)}\left( 
 \begin{array}{cc} q_{(1)} & 0 \end{array}
\right) U_{q(1)}^{\dagger}, 
\nonumber \\
\langle 0|\bar{q} |0\rangle &=& U_{\bar{q}(1)} \left(
 \begin{array}{c} \bar{q}_{(1)} \\ 0 \end{array}
\right) U_{c(2)}^{\dagger},  \\
\langle 0|T|0 \rangle     &=& U_{q(2)} T_{(1)} U_{\bar{q}(2)}^{\dagger},
\nonumber
\end{eqnarray}
where
\begin{equation}
U_{c(i)} \in SU(\tilde{N}_c), \qquad
U_{q(i)} \in SU(N_f)_Q , \qquad 
U_{\bar{q}(i)} \in SU(N_f)_{\bar Q},
\end{equation}
\begin{equation}
q_{(1)}=q_i \delta_i^j \qquad \bar{q}_{(1)}=\bar{q}_i \delta_i^j \qquad 
T_{(1)}=T_i \delta_i^j
\end{equation}
Without loss of generality, the values, $q_i$, $\bar{q}_i$ and $T_i$, 
$i=1,\cdots \tilde{N}_c$, can be chosen real and positive.

In this representation, the potential becomes
\begin{eqnarray}
  V(q,\bar{q},T) 
&\!\!\!
=&\!\!\! 
\frac1{k_T} {\rm tr}
(q_{(1)}^2 U_c \bar{q}_{(1)}^2 U_c^{\dagger}) 
+ \frac1{k_q} {\rm tr} 
 \left[ \left(
\begin{array}{cc}
q_{(1)}^2 & 0 \\
0 & 0
\end{array}
\right)
U_q^{\dagger} T_{(1)}^2 U_q +
U_{\bar{q}}^{\dagger} T_{(1)}^2 U_{\bar{q}} 
\left(
\begin{array}{cc}
\bar{q}_{(1)}^2 & 0 \\
0 & 0
\end{array}
\right)
 \right]
\nonumber \\
&\!\!\!+&\!\!\! 
\frac{\tilde g^2}2 
\left\{ {\rm tr}(q_{(1)}^2 \tilde t^a - U_c \bar{q}_{(1)}^2
    U_c^{\dagger} \tilde t^a) \right\}^2  \\
&+& {\rm tr} \left( U_{q(1)}^{\dagger} M_q^2 U_{q(1)} \left(
    \begin{array}{cc} q_{(1)}^2 & 0 \\ 0 & 0 \end{array} \right) +
      m_{\bar{q}}^2 \bar{q}_{(1)}^2 + m_T^2 T_{(1)}^2
\right), \nonumber
\end{eqnarray}
where $U_c = U_{c(1)}^{\dagger} U_{c(2)}$, $U_q = U_{q(2)}^{\dagger}
U_{q(1)}$, and $U_{\bar q} = U_{\bar{q}(2)}^{\dagger} U_{\bar{q}(1)}$.

We would like to find a minima of $V$ as a function of $q_i$, 
$\bar{q}_i$, $T_i$ and the unitary matrices $U$'s.
We notice immediately that the minimum of $V$ is at $T=0$
for fixed $q$, $\bar{q}$ and unitary matrices.
%
Varying next, at $T=0$, with respect to $U_c$ and $U_{q(1)}$,
we obtain 
\begin{eqnarray}
[q_{(1)}^2, U_c \bar{q}_{(1)}^2 U_c^{\dagger}] &=& 0, \\
\left[ U_{q(1)}^{\dagger} M_q^2 U_{q(1)}, \left(
    \begin{array}{cc} q_{(1)}^2 & 0 \\ 0 & 0 \end{array} 
    \right) \right] &=& 0.
\label{constraint}
\end{eqnarray}
These equations are satisfied only when
$U_c \bar{q}_{(1)}^2 U_c^{\dagger}$ 
and 
$U_{q(1)}^{\dagger} M_q^2 U_{q(1)}$ are diagonal matrices,
since $q_{(1)}^2$ is assumed to be a generic  diagonal matrix. 
The argument 
parallels the one given in the discussion of (\ref{eq:fourteen}). 
(The constraint (\ref{constraint}) allows the 
matrix $U_{q(1)}^{\dagger} M_q^2 U_{q(1)}$ to have 
arbitrary components in the last $N_f -\tilde N_c$ columns and rows.  
However, we can use the residual flavor symmetry to diagonalize the 
matrix to obtain the fully diagonal form.)

Since we assume that $n$ eigenvalues of $M^2_q$ is negative, 
we obtain the reduced potential
\begin{eqnarray}
V(q_i,\bar{q}_i) 
&=& \frac1{k_T} \sum_{i=1}^{\tilde{N}_c} q_i^2 \bar{q}_i^2 
+ \frac{\tilde g^2}{4 \tilde{N}_c} 
\sum_{i<j}^{\tilde{N}_c} (q_i^2 - \bar{q}_i^2
                                       -q_j^2 + \bar{q}_j^2)^2 \\
&-& m_{q_1}^2 \sum_{i=1}^n q_i^2 + m_{q_2}^2 \sum_{i=n+1}^{\tilde{N}_c} 
q_i^2   + m_{\bar q}^2 \sum_{i=1}^{\tilde{N}_c} \bar{q}_i^2,
\end{eqnarray}
up to permutations of  $q_i$ and $\bar{q}_i$.

{}For arbitrary values of the parameters, this potential is not bounded 
from below due to the negative sign of the term involving $m_{q_{1}}^2$. 
In order to keep the potential bounded from below, the soft-breaking 
masses must satisfy a condition, which is easily established by 
considering the special configuration $q_i=q$, $\bar{q}_i=0$ : 
\begin{equation}
-n m_{q_1}^2 + (\tilde{N}_c-n) m_{q_2}^2 \geq 0.
\label{bound}
\end{equation}

The extremum  conditions for the potential are given by 
\begin{eqnarray}
q_i \left(\frac1{k_T} \bar{q}_i^2 + \frac{\tilde g^2}{2 \tilde{N}_c}
\{ (\tilde{N}_c -1) (q_i^2-\bar{q}_i^2) - \sum_{k \neq i} (
q_k^2-\bar{q}_k^2) \} - m_{q_1}^2 \right) &=& 0, 
\label{eq:stationary1} \\
(i = 1,&\cdots,& n), \nonumber \\
q_i \left(\frac1{k_T} \bar{q}_i^2 + \frac{\tilde g^2}{2 \tilde{N}_c}
\{ (\tilde{N}_c -1) (q_i^2-\bar{q}_i^2) - \sum_{k\neq i} (
q_k^2-\bar{q}_k^2) \} + m_{q_2}^2 \right) &=& 0, 
\label{eq:stationary2} \\
(i = n+1 ,&\cdots,& \tilde{N}_c), \nonumber \\
\bar{q}_i \left(\frac1{k_T} q_i^2 - \frac{\tilde g^2}{2 \tilde{N}_c}
\{ (\tilde{N}_c -1) (q_i^2-\bar{q}_i^2) - \sum_{k\neq i} (
q_k^2-\bar{q}_k^2) \} + m_{\bar q}^2 \right) &=& 0, 
\label{eq:stationary3} \\
(i = 1,&\cdots,& \tilde{N}_c). \nonumber
\end{eqnarray}

To solve these equations, we first establish the following facts.

\begin{enumerate}
\item
{}For all $i$, we must have $q_i^2 \geq \bar{q}_i^2$. Indeed, suppose 
that there exist indices $i_1,\cdots, i_p$ for which 
$q_{i_{j}}^2 < \bar{q}_{i_{j}}^2$, while for all remaining indices 
$i_{k}$ we have $q_{i_{k}}^2 \geq\bar{q}_{i_{k}}^2$.  
This implies $\bar{q}_{i_j} \neq 0$, so that the expression
inside the  braces in  eq.~(\ref{eq:stationary3}) has to vanish for 
$i=i_1, \cdots, i_p$. Adding those equations, we obtain 
\begin{equation}
\frac1{k_T} \sum_{j=1}^p q_{i_j}^2 - \frac{\tilde g^2}{2 \tilde{N}_c}
\{ (\tilde{N}_c-p) \sum_{j=1}^p (q_{i_j}^2 - \bar{q}_{i_j}^2)
- p \sum_{k \neq i_1, \cdots, i_p} (q_k^2 -\bar{q}_k^2) \}
+ p m_{\bar q}^2 =0.
\end{equation}
Clearly, the left-hand-side is always positive and the equation has no
solutions (except for the solution $q=\bar q=0 $, only when 
$m_{\bar q} ^2 =0$). 

\item For all $i > n$, either $q_i=0$ or $\bar{q}_i=0$. 
Indeed, suppose that both $q_i\not=0$ and $\bar q_i\not=0$ for some $i$; 
then
by adding their respective equations in (\ref{eq:stationary2}) and
(\ref{eq:stationary3}), one obtains a left-hand-side which is strictly
positive, precluding the existence of any solution. Combining this point 
with
the one above in 1., it follows that $\bar q_i=0$ for all $i > n$.

\item For all $i > n$, $q_i=0$ as well. Indeed, if $q_1= \cdots =q_n =0$,
it follows from (\ref{eq:stationary2}) that $q_i$ are equal to zero  for 
all $i$.  If $q_1^2+\cdots +q_n^2\not=0$, this is proved  similarly to 1 
applying the stability condition  (\ref{bound}) to the sum of the 
expressions in braces in 
eq.~(\ref{eq:stationary1}) and eq.~(\ref{eq:stationary2}).

\end{enumerate}

{}From these facts, we find that the possible solutions of 
eq.~(\ref{eq:stationary1})-(\ref{eq:stationary3}) are 
\begin{equation}
q_1, \cdots, q_r \neq 0, \qquad \bar{q}_1, \cdots, \bar{q}_s \neq 0,
\qquad (n \geq r \geq s \geq 0),
\end{equation}
and the others are equal to zero,
up to permutations of $q_i$ and $\bar{q}_i$.
They dose not necessarily make minimal.

\begin{enumerate}

\item
The $r$ should be equal to $n$ at the minimal point.
Indeed, for $n \geq i > r$, the second derivative of $q_i$ is negative,
\begin{equation}
\frac{\partial^2 V}{\partial q_i^2}=
-\frac{\tilde{g}^2}{2 \tilde{N}_c} \sum_{k \neq i} (q_k^2-\bar{q}_k^2)
-m_{q_1}^2.
\end{equation}

\item
The $s$ should be equal to $n$ or 0 at the minimal point.
Indeed, for $n \geq i > s > 0$, the second derivative of $\bar{q}_i$ is
negative on the extremum solution.

\end{enumerate}

After all, we find that there are only two solutions for possible 
minimum,
described as follows. 
\begin{enumerate}
\item Only  $q_1, \cdots, q_n \neq 0$, while $q_i =0, ~i > n$ 
and $\bar q_i=0$ for all $i$.
The values of $q_1, \cdots, q_n$ are the same. We call the common value 
as $q_0$.
The value of $q_0$ and the potential in this configuration are
given by
\begin{equation}
q_0^2 = \frac2{\tilde g^2 \tilde{\gamma}} m_{q_1}^2,
\qquad \qquad
V= - \frac{n}{\tilde g^2 \tilde{\gamma}} m_{q_1}^4.
\end{equation}
where
\begin{equation}
\tilde{\gamma} = \frac{\tilde{N}_c-n}{\tilde{N}_c}.
\end{equation}

\item Only $q_1, \cdots, q_n \neq 0$ and $\bar{q}_1, \cdots \bar{q}_n 
\neq 0$, while  $q_i=\bar q_i=0, ~i > n$. 
The values of $q_0 \equiv q_1=\cdots=q_n$ and $\bar q_0\equiv \bar{q}_1=
\cdots \bar{q}_n$ are then given by
\begin{equation}
\left( \begin{array}{c} q_0^2 \\  \\ \bar{q}_0^2 \end{array} \right)
= \frac{1}{\tilde{\gamma} \tilde g^2 - \frac1{k_T}}
\left( \begin{array}{c}
\frac12 \tilde{\gamma} \tilde g^2k_T (m_{q_1}^2- m_{\bar q}^2)
        +  m_{\bar q}^2 \\ \\
\frac12 \tilde{\gamma} \tilde g^2k_T (m_{q_1}^2- m_{\bar q}^2)
        -  m_{q_1}^2 \end{array}
\right)
\label{vevofq}
\end{equation}
Given the fact that $q_1^2$ and $\bar q_1^2$ must be positive, this
expression yields a solution only when the following condition is 
satisfied
\begin{equation}
\frac12 \tilde{\gamma} \tilde g^2 k_T (m_{q_1}^2- m_{\bar q}^2) \geq  
m_{q_1}^2.
\label{eq:threethirty} 
\end{equation}
The value of the potential at the stationary point is given by 
\begin{equation}
V=- \frac{n}4 \frac{\tilde{\gamma} k_T \tilde g^2}
         {\tilde{\gamma} \tilde g^2 - \frac1{k_T}}
    \left( m_{\bar q}^2 - \frac{\frac12 \tilde{\gamma} \tilde g^2 
               - \frac1{k_T}}{\frac12 \tilde{\gamma} \tilde g^2 }
      m_{q_1}^2 \right)^2 -\frac{n}{ \tilde{\gamma}\tilde g^2}
      m_{q_1}^4
\end{equation}

\end{enumerate}

Therefore, whenever conditions (\ref{eq:threethirty}) is satisfied, 
solution 2 is the absolute minimum of the potential and describes the 
true ground state.  If condition (\ref{eq:threethirty}) is not 
satisfied, solution 1 is the absolute minimum.

\vspace{5mm}
\pagebreak[3]
\addtocounter{section}{0}
\addtocounter{subsection}{1}
\setcounter{footnote}{0}
\begin{center}
{\bf \thesubsection. Mass Spectrum }
\end{center}
\nopagebreak
\medskip
\nopagebreak
\hspace{2mm}

In our model, the Lagrangian has a global 
$SU(N_f-n)_Q \times SU(n)_Q \times U(1)_Q
\times SU(N_f)_{\bar Q} \times U(1)_B \times U(1)_R$ 
symmetry.
When the coupling $\tilde g$ is too weak, the
condition (\ref{eq:threethirty}) is not satisfied, solution 1. is 
the absolute minimum, and flavor symmetry is not broken. 
On the other hand, when the gauge coupling is 
strong, the condition (\ref{eq:threethirty}) is satisfied. 
Therefore solution 2 is the absolute minimum, and flavor symmetry is 
spontaneously broken as follows;
\begin{eqnarray}
&&SU(N_f-n)_Q \times SU(n)_Q \times U(1)_Q
\times SU(N_f)_{\bar Q} \times U(1)_B \times U(1)_R 
\\
&& \qquad 
\longrightarrow SU(N_f-n)_Q \times SU(N_f-n)_{\bar Q}
\times SU(n)_V \times U(1)_V 
\times U(1)_{B'} \times  U(1)_{R'}, \nonumber
\end{eqnarray}
where $SU(n)_V$ is the diagonal subgroup of $SU(n) \subset
SU(\tilde{N}_c)$ and $SU(n)_Q \times SU(n)_{\bar Q}$.
The spontaneous breaking of the global symmetry induces spontaneous 
breaking of color gauge symmetry  
$SU(\tilde{N}_c) \to SU(\tilde{N}_c-n)$.

Here we examine the mass spectrum after this spontaneous gauge symmetry 
breaking,  $SU(\tilde{N}_c) \to SU(n) \times SU(\tilde{N}_c-n)$.
To do so, we fix the gauge as follows,
\begin{equation}
\langle 0| q |0 \rangle = 
\left( \begin{array}{cc} q_0 I_n  & 0  \\
                            0     & 0  
       \end{array} \right)
\qquad
\langle 0| {\bar q} |0 \rangle = 
\left( \begin{array}{cc} \bar{q}_0 I_n & 0  \\
                                0      & 0   
       \end{array} \right)
\end{equation}
and we separate the field variables as follow,
$$
q  = 
\left ( \begin{array}{cc} 
     (q_0 + \frac{q_{\langle 0 \rangle}}{\sqrt n}) I_n
         + q_{\langle 1 \rangle} &  q_{\langle 2 \rangle}  \\
           q_{\langle 3 \rangle} &  q_{\langle 4 \rangle}  
        \end{array}
\right )\; , 
\qquad \qquad 
\psi_q  = 
\left ( \begin{array}{cc} 
           \frac{\psi_{\langle 0 \rangle}}{\sqrt n} I_n +
           \psi_{\langle 1 \rangle} &  \psi_{\langle 2 \rangle}  \\
           \psi_{\langle 3 \rangle} &  \psi_{\langle 4 \rangle}  
        \end{array}
\right )\; , 
$$
\begin{equation}
\bar q  = 
\left ( \begin{array}{cc} 
 (\bar{q}_0 + \frac{\bar{q}_{\langle \bar 0 \rangle}}{\sqrt n}) I_n
+\bar{q}_{\langle \bar 1 \rangle} & \bar{q}_{\langle \bar 3 \rangle} \\
 \bar{q}_{\langle \bar 2 \rangle} & \bar{q}_{\langle \bar 4 \rangle} 
        \end{array}
\right ),
\qquad \qquad
\psi_{\bar q}  = 
\left ( \begin{array}{cc} 
 \frac{\psi_{\langle \bar 0 \rangle}}{\sqrt n} I_n +
 \psi_{\langle \bar 1 \rangle} & \psi_{\langle \bar 3 \rangle} \\
 \psi_{\langle \bar 2 \rangle} & \psi_{\langle \bar 4 \rangle} 
        \end{array}
\right ),
\label{separation}
\end{equation}
$$
T =
\left ( \begin{array}{cc} 
     \frac{T_{\langle 0 \rangle}}{\sqrt n} I_n +
           T_{\langle 1 \rangle} &  T_{\langle 2 \rangle}  \\
           T_{\langle 3 \rangle} &  T_{\langle 4 \rangle}  \\
        \end{array}
\right ),
\qquad \qquad \qquad
\psi_T =
\left ( \begin{array}{cc} 
 \frac{\psi_{T \langle 0 \rangle}}{\sqrt n} I_n +
           \psi_{T \langle 1 \rangle} &  \psi_{T \langle 2 \rangle}  \\
            \psi_{T \langle 3 \rangle} &  \psi_{T \langle 4 \rangle}  
        \end{array}
\right ).  
$$
The gauge and gaugino fields can be similarly decomposed,
\begin{equation}
A_{\mu }   = 
\left ( \begin{array}{cc} 
     A_{\mu \langle 1 \rangle}           & A_{\mu \langle 2 \rangle} \\
     A_{\mu \langle 2 \rangle}^{\dagger} & A_{\mu \langle 3 \rangle} \\
        \end{array}
\right ) + A_{\mu \langle 0 \rangle} H_{\langle 0 \rangle}  
\qquad \quad 
\lambda   = 
\left ( \begin{array}{cc} 
  \lambda _{ \langle 1 \rangle }      & \lambda _{ \langle 2 \rangle } \\
   \lambda _{ \langle \bar 2 \rangle } & \lambda _{ \langle 3 \rangle} \\
        \end{array}
\right ) + \lambda _{ \langle 0 \rangle } H_{\langle 0 \rangle }   
\end{equation}
where 
\begin{equation}
H_{\langle 0 \rangle}=
\left ( \begin{array}{cc} 
              C_1 I_n    & 0 \\
              0      & -C_2 I_{\tilde{N}_c-n} 
        \end{array}
\right ),
\end{equation}
with 
$$ 
C_1 = \sqrt{\frac{\tilde{N}_c-n}{2 n \tilde{N}_c}}, \qquad   
   C_2 = \sqrt{\frac{n}{2  \tilde{N}_c (\tilde{N}_c -n)}}.
$$
The fields $q_{\langle 0 \rangle},~\bar{q}_{\langle 0 \rangle}$,
are uniquely defined, provided we insist on the following
conditions
\begin{eqnarray}
0&=&
{\rm \, tr}\, q_{\langle 1 \rangle} 
= {\rm \, tr}\, \bar q_{\langle \bar 1 \rangle} 
= {\rm \, tr}\, T_{\langle 1 \rangle} 
= {\rm \, tr} \psi _{\langle 1 \rangle} 
= {\rm \, tr} \psi _{\langle \bar 1 \rangle} 
= {\rm \, tr} \psi _{T\langle 1 \rangle}
\\
\nonumber
0&=&
{\rm \, tr} A_{\mu \langle 1 \rangle } 
= {\rm \, tr} A_{\mu \langle 3 \rangle} 
= {\rm \, tr} \lambda _{\langle 1 \rangle} 
= {\rm \, tr} \lambda _{\langle 3 \rangle}
\end{eqnarray}

\subsubsection{Spin 1 Masses}

Gauge boson mass terms can be obtained from the $D$-term part of the 
Lagrangian density, and given by
\begin{eqnarray}
&&- k_q \tilde{g}^2 
{\rm \, tr} \bigl \{ \langle0| q |0\rangle^\dagger A_\mu A^\mu  
\langle 0|q |0\rangle
           + \langle0| \bar{q} |0\rangle A_\mu A^\mu 
                    \langle0| \bar{q} |0\rangle^\dagger
    \bigl \} \\
&=& - \frac12 k_q \tilde{g}^2 (q_0^2+\bar{q}_0^2) \tilde{\gamma}
         A_{\mu \langle 0 \rangle} A_{\langle 0 \rangle}^\mu
    - k_q \tilde{g}^2 (q_0^2+\bar{q}_0^2)         
     {\rm \, tr} (A_{\mu \langle 1 \rangle} A_{\langle 1 \rangle}^{\mu}+
         A_{\mu \langle 2 \rangle} A_{\langle 2 \rangle}^{\mu \dagger}).
\nonumber
\end{eqnarray}
It is straightforward to read off the masses of the gauge bosons, 
and they are summarized in the first table~\ref{table:4}. 
The would-be-Goldstone bosons that are eaten in order to give these 
vector bosons mass are identified by inspecting the
linear coupling of the scalar particles to the gauge bosons. 
These couplings are given by
\begin{eqnarray}
&-&  k_q \tilde{g} \sqrt{2 \tilde{\gamma}}
   A_{\langle 0 \rangle}^\mu \partial_\mu 
   \left\{ {\rm Im} (q_0 q_{\langle 0 \rangle} 
              + \bar{q}_0 \bar{q}_{\langle \bar 0 \rangle}^*) \right\}
- 2 k_q \tilde{g} 
   {\rm \, tr} A_{\langle 1 \rangle}^\mu \partial_\mu 
   \left\{ {\rm Im} (q_0 q_{\langle 1 \rangle} 
        + \bar{q}_0 \bar{q}_{\langle \bar 1 \rangle}^\dagger) \right\}
\nonumber \\
& + & \left[i k_q \tilde{g} {\rm \, tr} 
   A_{\langle 2 \rangle}^\mu \partial_\mu 
   (q_0 q_{\langle 3 \rangle} 
              + \bar{q}_0 \bar{q}_{\langle \bar 3 \rangle}^\dagger)
 +  {\rm h.c.} \right]. 
\end{eqnarray}
The boson field  ${\rm Im} (q_0 q_{\langle 0 \rangle} 
+ \bar{q}_0 \bar{q}_{\langle \bar 0 \rangle}^*)$,
is eaten by the gauge field $A_{\mu \langle 0 \rangle}$, while the 
boson field, ${\rm Im} (q_0 q_{\langle 1 \rangle} 
+ \bar{q}_0 \bar{q}_{\langle \bar 1 \rangle}^\dagger)$,
are eaten by the gauge field, $A_{\mu \langle 1 \rangle}$, 
and the boson field, 
$(q_0 q_{\langle 3 \rangle} 
+ \bar{q}_0 \bar{q}_{\langle \bar 3 \rangle}^\dagger)$,
(and its complex conjugate) are eaten by the gauge field, 
$A_{\mu \langle 2 \rangle}$.

The properties of the vector bosons are summarized in the 
table~\ref{table:4}.

\begin{table}[t]
\begin{center}
\begin{tabular}{|c|c|c|c|} \hline 
Spin $1$  & Masses$^2$ & Multiplicity  & $SU(\tilde{N}_c-n)$
                \\ \hline \hline
$A_{\mu \langle 0 \rangle}$ & $
           k_q \tilde \gamma \tilde g^2 (q_0^2+\bar{q}_0^2)$  & 1 & $1$
                \\ \hline
$A_{\mu \langle 1 \rangle}$ & $
                k_q \tilde g^2 (q_0^2+\bar{q}_0^2)$  & $n^2-1$ 
                & 1 
                \\ \hline
$A_{\mu \langle 2 \rangle}$&$\frac12 k_q \tilde g^2 (q_0^2+\bar{q}_0^2)$ 
             & $2 n (\tilde{N}_c-n)$ & $(\tilde{N}_c-n)
                \oplus (\tilde{N}_c-n)^*$ 
                \\ \hline
$A_{\mu \langle 3 \rangle}$ & 0 
             & $ (\tilde{N}_c-n)^2 -1$ & adjoint
                \\ \hline
\end{tabular}
\end{center}
\caption{The spin 1 fields and their masses in the dual theory}
\label{table:4}
\end{table}

\subsubsection{Spin 1/2 Masses}
The fermion mass terms are governed by contributions both from the 
superpotential (contribution $V_1$) and from the mixing of the 
quarks and the gauginos in the $D$-term (contribution $V_2$) in 
eq.(\ref{eq:three}). The first contribution, in
terms of the irreducible fields of eq.(\ref{separation}) is 
given by
\begin{eqnarray}
V_1 &=& q_0  \{ \psi_{T \langle 0 \rangle} \psi_{\langle \bar 0 \rangle}
  +{\rm \, tr} ( \psi_{T \langle 1 \rangle} \psi_{\langle \bar 1 \rangle}
           +\psi_{T \langle 2 \rangle} \psi_{\langle \bar 2 \rangle})\}
\nonumber \\
&+& \bar{q}_0 \{ \psi_{\langle 0 \rangle} \psi_{T \langle 0 \rangle}
  +{\rm \, tr} ( \psi_{\langle 1 \rangle} \psi_{T \langle  1 \rangle}
           +\psi_{\langle 2 \rangle} \psi_{T \langle 3 \rangle}) \}
     +{\rm h.c.}
\end{eqnarray}
Similarly, the contribution $V_2$ may be decomposed in terms of
the irreducible fields of eq.(\ref{separation}),
and we obtain
\begin{eqnarray}
V_2 &=& - i \sqrt{2} \tilde g k_q {\rm \, tr} 
\langle 0| q |0\rangle^\dagger
                        \lambda \psi_q 
        + i \sqrt{2} \tilde g k_q {\rm \, tr} \psi_{\bar q} \lambda 
                    \langle 0| \bar q |0\rangle^\dagger + {\rm h.c.} \\
    &=& - i \tilde g k_q \sqrt{\tilde{\gamma}} 
         \left[ q_0 \lambda_{\langle 0 \rangle} \psi_{\langle 0 \rangle}
    -\bar{q}_0 \psi_{\langle \bar 0 \rangle} \lambda_{\langle 0 \rangle} 
           \right]
\nonumber \\
    &&  -i \sqrt{2} \tilde g k_q 
       \left[ q_0 {\rm \, tr} (\lambda_{\langle 1 \rangle} 
\psi_{\langle 1 \rangle}
              +\lambda_{\langle 2 \rangle} \psi_{\langle 3 \rangle})
       -\bar{q}_0 {\rm \, tr} (\psi_{\langle \bar 1 \rangle} 
\lambda_{\langle 1 \rangle} 
               + \psi_{\langle \bar 3 \rangle} \lambda_{\langle \bar 2 
\rangle})   \right]
           +{\rm h.c.} \nonumber 
\end{eqnarray}

{}From these mass terms, we can get the mass spectrum, but one must take 
into account the effective coupling normalizations of the kinetic terms 
for the spinors. To do so, it is convenient to normalize the fermion 
fields to canonical expressions for the kinetic terms by the following 
rescalings
\begin{equation}
\psi_q^\prime        = \sqrt{k_q} \psi_q, \qquad
\psi_{\bar q}^\prime = \sqrt{k_q} \psi_{\bar q}, \qquad
\psi_T^\prime        = \sqrt{k_T} \psi_T.
\end{equation}
The gaugino $\lambda$ already has canonical normalization by gauge 
invariance and need not be rescaled. 
The fields may be separated into groups transforming under different
irreducible representations of the color and flavor groups. 
Singlets under color symmetry $SU(\tilde{N}_c-n)$ are
$(\psi_{\langle 2 \rangle},   \psi_{T \langle 3 \rangle})$,
$(\psi_{T \langle 2 \rangle}, \psi_{\langle \bar 2 \rangle})$,
$( \lambda_{\langle 0 \rangle},
              \psi_{\langle 0 \rangle},
              \psi_{\langle \bar 0 \rangle},
              \psi_{T \langle 0 \rangle} )$,
and        $( \lambda_{\langle 1 \rangle},
              \psi_{\langle 1 \rangle},
              \psi_{\langle \bar 1 \rangle},
              \psi_{T \langle 1 \rangle} )$,
and have Majorana masses respectively.
{}Fundamental under $SU(\tilde{N}_c-n)$ are
$(\lambda_{\langle 2 \rangle}, \psi_{\langle \bar 3 \rangle})$ and
anti-fundamental are
$(\lambda_{\langle \bar 2 \rangle}, \psi_{\langle 3 \rangle})$ which have
Dirac masses between each other.

In the set $( \lambda_{\langle 0 \rangle},
              \psi_{\langle 0 \rangle},
              \psi_{\langle \bar 0 \rangle},
              \psi_{T \langle 0 \rangle} )$,
the masses are governed by the quartic equation
\begin{equation}
0 = {\cal M}^4 - ( \tilde{\gamma} \tilde g^2 k_q + \frac1{k_q k_T} )
                 ( q_0^2 + \bar{q}_0^2 ) {\cal M}^2
         + \tilde{\gamma} \frac1{k_T} \tilde g^2 (q_0^2 + \bar{q}_0^2)^2.
\end{equation}
The solutions are 
\begin{eqnarray}
{\cal M}^2_{\langle 0\rangle} &=&  k_q \tilde g^2 \tilde{\gamma}
                    (q_0^2 + \bar{q}_0^2) \nonumber \\
{\cal M}^2_{T \langle 0 \rangle} & =&          \frac1{k_q k_T} (q_0^2 +
\bar{q}_0^2).
\end{eqnarray}

In the set $( \lambda_{\langle 1 \rangle},
              \psi_{\langle 1 \rangle},
              \psi_{\langle \bar 1 \rangle},
              \psi_{T \langle 1 \rangle} )$,
the masses are governed by the quartic equation
\begin{equation}
0 = {\cal M}^4 - ( 2 \tilde g^2 k_q + \frac1{k_q k_T} )
                 ( q_0^2 + \bar{q}_0^2 ) {\cal M}^2
           +2 \frac1{k_T} \tilde g^2 (q_0^2 + \bar{q}_0^2)^2.
\end{equation}
The solutions are 
\begin{eqnarray}
{\cal M}^2_{\langle 1\rangle} &=&  2 k_q \tilde g^2 
                    (q_0^2 + \bar{q}_0^2) \nonumber \\
{\cal M}^2_{T\langle 1\rangle} & =&          \frac1{k_q k_T} (q_0^2 +
\bar{q}_0^2).
\end{eqnarray}

A summary of all spin 1/2 fields and masses is given in
the table~\ref{table:5}.

\begin{table}[t]
\begin{center}
\begin{tabular}{|c|c|c|c|} \hline 
Spin $1/2$  & Masses$^2$ & Multiplicity  & $SU(\tilde{N}_c-n)$
  \\ \hline \hline
$\psi_{\langle 2 \rangle},~\psi_{T \langle 3 \rangle}$
& $\frac1{k_q k_T} \bar{q}_0^2$
   & $2 n (N_f-n) $ & 1 
  \\ \hline
$\psi_{T \langle 2 \rangle},~\psi_{\langle \bar 2 \rangle}$
& $\frac1{k_q k_T} q_0^2$
   & $2 n (N_f-n) $ & 1
  \\ \hline
$ \lambda_{\langle 0 \rangle},~\psi_{T \langle 0 \rangle},$
& ${\cal M}^2_{\langle 0\rangle}$
   & 4 & 1 \\
$ \psi_{\langle 0 \rangle},~\psi_{\langle \bar 0 \rangle}$
& ${\cal M}^2_{T\langle 0\rangle}$ 
   &   &   
  \\ \hline
$ \lambda_{\langle 1 \rangle},~\psi_{T \langle 1 \rangle},$
& ${\cal M}^2_{\langle 1\rangle}$
   & $4 (n^2-1)$ & 1 \\
$ \psi_{\langle 1 \rangle},~\psi_{\langle \bar 1 \rangle}$
& ${\cal M}^2_{T\langle 1\rangle}$ 
   &   &   
  \\ \hline
$(\lambda_{\langle 2 \rangle},~\psi_{\langle \bar 3 \rangle})$ 
   &
   & $2 n (\tilde{N}_c-n)$ &  $\tilde{N}_c-n$ \\
      & $2 \tilde g^2 k_q q_1^2, 2 \tilde g^2 k_q \bar{q}_1^2$ & & \\
$(\lambda_{\langle \bar 2 \rangle},~\psi_{\langle 3 \rangle})$ 
   &   & $2 n (\tilde{N}_c-n)$ &  $(\tilde{N}_c-n)^*$ 
  \\ \hline
\end{tabular}
\end{center}
\caption{The spin 1/2 fields and their masses in the dual theory}
\label{table:5}
\end{table}

\subsubsection{Spin 0 Masses}

We give the mass terms of spin 0 masses in terms of the irreducible
fields of eq.(\ref{separation}).

In the case of weak coupling, the mass terms are as follows,
\begin{eqnarray}
- {\cal L}_{\rm mass} &\!\!\!=&\!\!\! 
    m_{q_1}^2 \left(
          q_{\langle 0 \rangle}+q_{\langle 0 \rangle}^*
              \right)^2 +
    \frac1{\tilde{\gamma}} m_{q_1}^2 {\rm \, tr} \left(
          q_{\langle 1 \rangle}+q_{\langle 1 \rangle}^\dagger
              \right)^2 \nonumber \\
&\!\!\!+&\!\!\! 
(m_{q_1}^2 + m_{q_2}^2) 
        {\rm \, tr} q_{\langle 2 \rangle} q_{\langle 2 \rangle}^\dagger +
   (m_{q_2}^2-\frac{n}{\tilde{N}_c-n} m_{q_1}^2) 
   {\rm \, tr} q_{\langle 4 \rangle} 
q_{\langle 4 \rangle}^\dagger \nonumber \\
&\!\!\!+&\!\!\! \left\{ \left( \frac2{\tilde g^2 \tilde{\gamma}}
                   \frac1{k_T}-1 \right) m_{q_1}^2 + m_{\bar q}^2
    \right\} (\bar{q}_{\langle \bar 0 \rangle}^*
              \bar{q}_{\langle \bar 0 \rangle} +
          {\rm \, tr} \bar{q}_{\langle \bar 1 \rangle}^\dagger
              \bar{q}_{\langle \bar 1 \rangle} +
          {\rm \, tr} \bar{q}_{\langle \bar 2 \rangle}^\dagger 
              \bar{q}_{\langle \bar 2 \rangle}) \nonumber \\
&\!\!\!+&\!\!\!
    \left( m_{\bar q}^2 + \frac{n}{\tilde{N}_c-n} m_{q_1}^2 \right)
    {\rm \, tr} ( \bar{q}_{\langle \bar 3 \rangle}^\dagger 
         \bar{q}_{\langle \bar 3 \rangle} +
         \bar{q}_{\langle \bar 4 \rangle}^\dagger 
         \bar{q}_{\langle \bar 4 \rangle}) \\
&\!\!\!+&\!\!\!
 (\frac1{k_q} \frac2{\tilde g^2 \tilde{\gamma}} m_{q_1}^2 + m_T^2)
     (T_{\langle 0 \rangle} T_{\langle 0 \rangle}^* 
    + {\rm \, tr} T_{\langle 1 \rangle} T_{\langle 1 \rangle}^\dagger 
    + {\rm \, tr} T_{\langle 2 \rangle} T_{\langle 2 \rangle}^\dagger ) 
\nonumber \\
&\!\!\!+&\!\!\!
  m_T^2  {\rm \, tr} (T_{\langle 3 \rangle} T_{\langle 3 \rangle}^\dagger
            + T_{\langle 4 \rangle} T_{\langle 4 \rangle}^\dagger )
\nonumber
\end{eqnarray}

The bosons, ${\rm Im}~q_{\langle 0 \rangle}$, 
$q_{\langle 1 \rangle}- q_{\langle 1 \rangle}^\dagger$, 
$q_{\langle 3 \rangle}$ and
$q_{\langle 3 \rangle}^\dagger$ will be eaten by gauge bosons.
There is no genuine Nambu-Goldstone boson in this case.

In the case of strong coupling, the mass spectrum is as follows;
\begin{eqnarray}
- {\cal L}_{\rm mass} &\!\!\!=&\!\!\! 
\frac14 \tilde{\gamma} \tilde g^2 q_0^2 \left(
          q_{\langle 0 \rangle}+q_{\langle 0 \rangle}^*
              \right)^2 +
\frac14 \tilde{\gamma} \tilde g^2 \bar{q}_0^2 \left(
          \bar{q}_{\langle \bar 0 \rangle}
               +\bar{q}_{\langle \bar 0 \rangle}^*
              \right)^2      \nonumber \\
&\!\!\!-&\!\!\!
  (\frac12 \tilde{\gamma} \tilde g^2 - \frac1{k_T}) q_0 \bar{q}_0 
     \left(q_{\langle 0 \rangle}+q_{\langle 0 \rangle}^*
     \right)
     \left(\bar{q}_{\langle \bar 0 \rangle}
                +\bar{q}_{\langle \bar 0 \rangle}^*
     \right) \nonumber \\
&\!\!\!+&\!\!\!
\frac14  \tilde g^2 q_0^2 {\rm \, tr} \left(
          q_{\langle 1 \rangle}+q_{\langle 1 \rangle}^\dagger
              \right)^2 +
\frac14  \tilde g^2 \bar{q}_0^2 \left(
          \bar{q}_{\langle \bar 1 \rangle}
               +\bar{q}_{\langle \bar 1 \rangle}^\dagger
              \right)^2      \nonumber \\
&\!\!\!-&\!\!\!\!
 (\frac12 \tilde g^2 - \frac1{k_T}) q_0 \bar{q}_0 {\rm \, tr}
     \left(q_{\langle 1 \rangle}+q_{\langle 1 \rangle}^\dagger
     \right)
     \left(\bar{q}_{\langle \bar 1 \rangle}
                +\bar{q}_{\langle \bar 1 \rangle}^\dagger
     \right) \nonumber \\
&\!\!\!+&\!\!\!
(m_{q_1}^2 + m_{q_2}^2) {\rm \, tr} q_{\langle 2 \rangle}
q_{\langle 2 \rangle}^\dagger 
+ \left( \frac12 \tilde g^2 k_T - 1 \right) 
(m_{q_1}^2
    - m_{\bar q}^2)  \left|
      \frac{ \bar{q}_0 q_{\langle 3 \rangle}
            - q_0 \bar{q}_{\langle \bar 3 \rangle}^\dagger }
           {\sqrt{q_0^2+ 
                  \bar{q}_0^2}} \right|^2 \nonumber \\
&\!\!\!+&\!\!\!\ \left(m_{q_2}^2 - \frac{n}{\tilde{N}_c-n} 
                     \frac{\frac12 \tilde{\gamma} \tilde g^2}
                    {\tilde{\gamma} \tilde g^2 - \frac1{k_T}}
         (m_{q_1}^2 + m_{\bar q}^2) \right)
    {\rm \, tr} q_{\langle 4 \rangle} q_{\langle 4 \rangle}^\dagger
    \nonumber \\
&\!\!\!+&\!\!\! \left(m_{\bar q}^2 + \frac{n}{\tilde{N}_c-n} 
                     \frac{\frac12 \tilde{\gamma} \tilde g^2}
                    {\tilde{\gamma} \tilde g^2 - \frac1{k_T}}
         (m_{q_1}^2 + m_{\bar q}^2) \right)
    {\rm \, tr} \bar{q}_{\langle \bar 4 \rangle}^\dagger
               \bar{q}_{\langle \bar 4 \rangle}
    \nonumber \\
&\!\!\!+&\!\!\!
 \left\{\frac{k_T}{k_q} (m_{q_1}^2 - m_{\bar q}^2) + m_T^2 \right\}
 (T_{\langle 0 \rangle} T_{\langle 0 \rangle}^* +
  {\rm \, tr} T_{\langle 1 \rangle} T_{\langle 1 \rangle}^\dagger)
    \\
&\!\!\!+&\!\!\! (\frac1{k_q} q_0^2 + m_T^2)
         {\rm \, tr} T_{\langle 2 \rangle} T_{\langle 2 \rangle}^\dagger
           + (\frac1{k_q} \bar{q}_0^2 + m_T^2)
          {\rm \, tr} T_{\langle 3 \rangle}^\dagger T_{\langle 3 \rangle}
  + m_T^2 {\rm \, tr} T_{\langle 4 \rangle}^\dagger T_{\langle 4 \rangle}
    \nonumber 
\end{eqnarray}

The bosons, ${\rm Re} q_{\langle 1 \rangle}$ and 
${\rm Re} \bar{q}_{\langle \bar 1 \rangle}$, will mix.
The mass eigenvalues are governed by the equation
\begin{equation}
0= ({\cal M}^2)^2 - \frac1{4 k_q} \tilde{\gamma} \tilde g^2 
(q_0^2+\bar{q}_0^2)
                 {\cal M}^2
    +\frac1{4 k_T k_q} (\tilde{\gamma} \tilde g^2 -\frac1{k_T}) 
q_0^2 \bar{q}_0
\end{equation}
The solutions in the limit where $\tilde g^2 \gg 1$ are
\begin{equation}
{\cal M}^2 = 
\tilde{\gamma} \tilde g^2 \frac{k_T}{k_q} (m_{q_1}^2-m_{\bar q}^2),
\qquad m_{q_1}^2-m_{\bar q}^2.
\end{equation}

As shown in previous section,
the bosons, ${\rm Im} (q_0 q_{\langle 0 \rangle} 
              + \bar{q}_0 \bar{q}_{\langle \bar 0 \rangle}^*)$,
 ${\rm Im} (q_0 q_{\langle 1 \rangle} 
              + \bar{q}_0 \bar{q}_{\langle \bar 1 \rangle}^\dagger)$,
$(q_0 q_{\langle 3 \rangle} 
              + \bar{q}_0 \bar{q}_{\langle \bar 3 \rangle}^\dagger)$,
and its complex conjugates are would-be-Goldstone bosons  
which are eaten by gauge bosons.
The bosons, ${\rm Im} (\bar{q}_0 q_{\langle 0 \rangle} 
              - q_0 \bar{q}_{\langle \bar 0 \rangle}^*)$,
${\rm Im} (\bar{q}_0 q_{\langle 1 \rangle} 
              - q_0 \bar{q}_{\langle \bar 1 \rangle}^\dagger)$,
$\bar{q}_{\langle \bar 2 \rangle}$ and 
$\bar{q}_{\langle \bar 2 \rangle}^\dagger$ 
are genuine Nambu-Goldstone bosons corresponding to
$SU(N_f) \rightarrow SU(N_f-n)_{\bar Q} \times
SU(n)_V$.
%
\vspace{7mm}
\pagebreak[3]
\addtocounter{section}{1}
\setcounter{equation}{0}
\setcounter{subsection}{0}
\setcounter{footnote}{0}
\begin{center}
{\large {\bf \thesection. Discussion 
on the validity of our analysis}}
\end{center}
\nopagebreak
\medskip
\nopagebreak
\hspace{3mm}
It has been found that the classical moduli space is modified 
quantum mechanically for cases $N_f=N_c$ 
\cite{AfDiSe}, \cite{Seiberg}. 
{}For $N_f \ge N_c$, we can define baryon fields 
\begin{equation}
B^{i_1, \cdots, i_{N_c}}=\epsilon_{a_1,\cdots, a_{N_c}} 
Q_{a_1}{}^{i_1} \cdots Q_{a_{N_c}}{}^{i_{N_c}} .
\end{equation}
In terms of the baryon and antibaryon field $B$ and $\bar B$
and meson field $T$ in eq.(\ref{eq:two}), 
there is a constraint classically for $N_f=N_c$ which is 
modified as 
\begin{equation}
{\rm det} T - B \bar B =0 \rightarrow 
{\rm det} T - B \bar B = \Lambda^{2N_c}
\end{equation}
In such a situation, the quantum fluctuations are important 
and these meson and baryon fields cannot be treated 
semi-classically as a product of elementary quark fields. 
In particular, we cannot regard the vacuum expectation value of 
the meson and baryon fields to be a product of those of the elementary 
squark fields. 
Therefore even the analysis of vacuum configuration requires 
more than a semi-classical treatment which is used in our paper. 
This phenomenon clearly suggests that the semi-classical treatment 
of these meson and baryon fields as a product of 
elementary quark fields is not accurate. 

{}For $N_f=N_c+1$, the classical moduli space is not 
modified \cite{AfDiSe}, \cite{Seiberg}.  
\begin{equation}
B_i T^i{}_j \bar B^j - {\rm det} T =0
\label{moduliconstraint}
\end{equation}
Therefore we can use the classical description for vacuum expectation 
values. 
On the other hand, the nonperturbative potential is given in terms of 
this combination of scalar fields  
\begin{equation}
W=B_i T^i{}_j \bar B^j - {\rm det} T 
\label{dualsuperpot}
\end{equation}
If the scalar meson and baryon fields are just the simple product 
of scalar quark fields,  the relation (\ref{moduliconstraint}) should 
hold for fields themselves. 
Therefore if we take 
fundamental fields $Q$ and $\bar Q$ as dynamical variables 
and regard the composite fields $T$, $B$, and $\bar B$ as 
merely the naive products of them, 
the superpotential vanishes identically.
Hence the superpotential (\ref{dualsuperpot}) implies 
that we cannot regard the meson and baryon fields to be 
the products of quark fields, and the quantum fluctuations are important 
for these composite fields. 

At the moment we have no good means of analyzing the case where 
quantum mechanical effects are really important such as $N_f=N_c$ and 
$N_f=N_c+1$. 
Our analysis should be reliable away from these situations 
($N_f \ll N_c$ or $N_f \gg N_c$) and hopefully carries a qualitatively 
correct result even for 
situations where such quantum effects are important. 

Let us finally reconsider the region of validity of our analysis. 
The most reliable parameter region is clearly the case of small 
soft mass terms (compared to the gauge interaction scale $\Lambda$). 
In order to avoid the vacuum instability in the limit of vanishing 
soft mass terms, we can add a small supersymmetric mass. 
If we do not have the soft mass terms and have only a generic 
supersymmetric mass together with the nonperturbative superpotential, 
we obtain a stable supersymmetric vacuum with vacuum expectation 
values for squarks which break gauge as well as flavor symmetries. 
We can introduce the soft mass terms to this system and 
repeat the analysis almost identical to the one in this paper. 
Then we find that the results 
smoothly connect to those of the supersymmetric vacuum 
in the limit of vanishing soft mass terms. 
The region near this situation is most reliable and is nearly 
perturbative. 
As we increase the value of soft mass terms, we may encounter 
modifications of the nonperturbative effects due to the soft mass 
terms. 
Assuming such modifications are small, we can explore an interesting 
situation where the gauge (and flavor) symmetries are broken 
mainly by the negative soft mass squared terms and calculate the 
nonperturbative corrections for vacuum expectation values and masses 
due to the (mildly strong) gauge interactions. 
This possibility is a genuinely new interesting case, although 
it clearly involves more assumptions. 
This assumption is partly supported by our analysis of soft breaking 
terms using the spurion in sect.~2.1. 
By the similar token, our analysis for the case of $N_f > N_c$ is 
accurate when we have small soft mass squared terms for dual 
quarks. 
However, the relation between dual quark field and the original 
fundamental quark field is nonperturbative and complicated. 
Therefore the relation between the soft mass terms of the original 
squark field and those of the dual quark fields may not be simple. 
This ambiguity is reflected in the fact that we have the unknown 
coefficient for the K\"ahler potential of the kinetic term for 
dual quarks $k_q$ and for mesons $k_T$ in eq.(\ref{kaehler}). 
In this respect, our analysis for the case $N_f > N_c$ involves more 
assumptions than that of $N_f < N_c$ case. 

\vspace{5mm} 
%
%
We acknowledge the hospitality of Institute for Theoretical Physics in 
Santa Barbara for providing the stimulating atmosphere in which most of 
this research was carried out. Also, Y.M and N.S. thank the Department 
of Physics at UCLA for their hospitality.  This work is partially 
supported by the US-Japan collaborative  program, by National Science 
Foundation Grant PHY-92-18990 and by Grant-in-Aid for  Scientific 
Research (Y.M.) and (No.05640334 for N.S.) 
from the Japanese Ministry of Education, Science and Culture.
\vspace{5mm} 
%
\vspace{5mm}
%
\newcommand{\NP}[1]{{\it Nucl.\ Phys.\ }{\bf #1}}
\newcommand{\PL}[1]{{\it Phys.\ Lett.\ }{\bf #1}}
\newcommand{\CMP}[1]{{\it Commun.\ Math.\ Phys.\ }{\bf #1}}
\newcommand{\MPL}[1]{{\it Mod.\ Phys.\ Lett.\ }{\bf #1}}
\newcommand{\IJMP}[1]{{\it Int.\ J. Mod.\ Phys.\ }{\bf #1}}
\newcommand{\PRP}[1]{{\it Phys.\ Rep.\ }{\bf #1}}
\newcommand{\PR}[1]{{\it Phys.\ Rev.\ }{\bf #1}}
\newcommand{\PRL}[1]{{\it Phys.\ Rev.\ Lett.\ }{\bf #1}}
\newcommand{\PTP}[1]{{\it Prog.\ Theor.\ Phys.\ }{\bf #1}}
\newcommand{\PTPS}[1]{{\it Prog.\ Theor.\ Phys.\ Suppl.\ }{\bf #1}}
\newcommand{\AP}[1]{{\it Ann.\ Phys.\ }{\bf #1}}
\newcommand{\ZP}[1]{{\it Zeit.\ f.\ Phys.\ }{\bf #1}}
\end{document}